\documentclass[aps,prd,10pt,floatfix,amsmath,amssymb,twocolumn,longbibliography,superscriptaddress,nofootinbib]{revtex4-2}

\usepackage[utf8]{inputenc}
\usepackage{lmodern}
\usepackage[T1]{fontenc}
\usepackage{verbatim} 
\usepackage{mathtools}
\usepackage{graphicx}
\usepackage[dvipsnames]{xcolor}
\usepackage{ragged2e}
\usepackage{subcaption} 
\DeclareCaptionJustification{justified}{\justifying}
\usepackage[toc,page]{appendix}
\usepackage{array} 
\usepackage[pdftex,breaklinks,colorlinks,
linkcolor=Blue,
citecolor=teal,
anchorcolor=red,
urlcolor=cyan,
pdfencoding=auto]{hyperref}
\usepackage{orcidlink}

\usepackage{url}
\usepackage{mathrsfs} 
\graphicspath{{Plots/}}

\newcommand{\Phiret}{\Phi^{\rm ret}}
\newcommand{\PhiR}{\Phi^{\rm R}}
\newcommand{\Phires}{\Phi^\mathcal{R}}
\newcommand{\PhiP}{\Phi^\mathcal{P}}
\newcommand{\PhiS}{\Phi^{\rm S}}

\newcommand{\xpart}{{x^\mu_p}}
\newcommand{\Plm}{P_\ell^m}
\newcommand{\Psibar}{\bar{\Psi}}
\newcommand{\cR}{{{R}}}
\newcommand{\cC}{{\bar{C}}}
\newcommand{\cl}{{\bar{\ell}}}
\newcommand{\cm}{{\bar{m}}}
\newcommand{\cn}{{{n}}}
\newcommand{\cpp}{{\bar{p}}}
\newcommand{\ctheta}{{\bar{\theta}}}
\newcommand{\cphi}{{\bar{\varphi}}}

\newcommand{\suml}{\sum_{\cl=0}^{\cl_\text{max}}}
\newcommand{\summ}{\sum_{\cm=-\cl}^\cl}

\newcommand{\sumnFull}{\sum_{\cn=-1}^{\cn_\text{max}}}
\newcommand{\Ylm}{Y_{\cl\cm}}
\newcommand{\Ymod}{\tilde{Y}_{\cl\cm}}
\newcommand{\Ilmnm}{\mathfrak{I}_{\cl \cm \cn}^m}

\newcommand{\bracket}[1]{\left( #1 \right)}

\newcommand{\floor}[1]{\left\lfloor #1 \right\rfloor}

\newcommand{\red}{\color{red}}

\newcommand{\AP}[1]{{\color{red}AP: #1}}
\newcommand{\PB}[1]{{\color{orange}PB: #1}}

\begin{document}


\title{Simple, efficient method of calculating the Detweiler-Whiting singular field to very high order}

\author{Patrick Bourg\,\orcidlink{0000-0003-0015-0861}}
\affiliation{School of Mathematical Sciences and STAG Research Centre, University of Southampton,
	Southampton SO17 1BJ, United Kingdom}
\affiliation{Institute for Mathematics, Astrophysics and Particle Physics, Radboud University, Heyendaalseweg 135, 6525 AJ Nijmegen, The Netherlands}
\author{Adam Pound\,\orcidlink{0000-0001-9446-0638}}
\affiliation{School of Mathematical Sciences and STAG Research Centre, University of Southampton,
	Southampton SO17 1BJ, United Kingdom}
\author{Samuel D.\ Upton\,\orcidlink{0000-0003-2965-7674}}
\affiliation{School of Mathematical Sciences and STAG Research Centre, University of Southampton,
	Southampton SO17 1BJ, United Kingdom}
\affiliation{Astronomical Institute of the Czech Academy of Sciences, Bo\v{c}n\'{i} II 1401/1a, CZ-141 00 Prague, Czech Republic}
\author{Rodrigo Panosso Macedo\,\orcidlink{0000-0003-2942-5080}}
\affiliation{Niels Bohr International Academy, Niels Bohr Institute, Blegdamsvej 17, 2100 Copenhagen, Denmark}
\date{\today}

\begin{abstract}
    Most self-force calculations rely, in one way or another, on representations of a particle's Detweiler-Whiting singular field. We present a simple method of calculating the singular field to high order in a local expansion in powers of distance from the particle. As a demonstration, we compute the singular field to 14th order in distance, 10 orders beyond the previous state of the art, in the simple case of a scalar charge in circular orbit around a Schwarzschild black hole. We provide the result in both a 4-dimensional form and a decomposed form suitable for use in an $m$-mode puncture scheme. Our method should have applications in overcoming bottlenecks in current self-force calculations at both first and second order in perturbation theory.
\end{abstract}

\maketitle

\tableofcontents

\section{Introduction}

Self-force theory is the primary approach to modeling compact binaries with small mass ratios~\cite{Barack:2018yvs,Pound:2021qin}. Recent progress in the method include (i) the creation of a software package for generating self-force-based gravitational waveforms sufficiently rapidly for data analysis~\cite{Chua:2020stf,Katz:2021yft}, (ii) inclusion of the smaller object's spin in a variety of binary configurations~\cite{Skoupy:2021asz,Mathews:2021rod,Drummond:2022xej,Skoupy:2023lih}, (iii) modelling of transient resonances~\cite{Gupta:2022fbe}, (iv) calculations of the scattering angle in hyperbolic encounters~\cite{Barack:2022pde,Whittall:2023xjp}, (v) calculation of the first-order gravitational self-force on generic, inclined and eccentric orbits around a Kerr black hole~\cite{vandeMeent:2017bcc}, (vi) rapid generation of waveforms with high eccentricity~\cite{McCart:2021upc}, (vii) generation of generic waveforms at leading, adiabatic order (0PA) in a two-timescale expansion~\cite{Hughes:2021exa,*Hughes2023,Isoyama:2021jjd}, and (viii) computation of waveforms at first post-adiabatic order  (1PA) in the restricted case of quasicircular, nonspinning binaries~\cite{Wardell:2021fyy}.

In the context of binaries, the gravitational self-force approach is based on an expansion of the spacetime metric in powers of the binary's mass ratio, $\varepsilon=m/M$. The zeroth-order term in this expansion is the metric of the large body of mass $M$, typically a Kerr black hole, and the order-$\varepsilon$ and higher-order terms are metric perturbations generated by the small companion of mass $m$, which satisfy the perturbative Einstein equations on the background spacetime.

Waveform generation in this approach is typically broken into two steps~\cite{Miller:2020bft,Hughes:2021exa,Katz:2021yft,Pound:2021qin}: an offline step and an online step. In the offline step, one solves the perturbative field equations on a grid of binary parameter values. At each point in the grid, one computes and stores waveform mode amplitudes and rates of change of orbital parameters. In the online step, one then solves simple ordinary differential equations to rapidly move through the parameter space and generate a waveform. Even for rapidly precessing, highly eccentric orbits, the online step is extremely fast. However, it relies on first completing the offline step. Currently, this offline step is computationally expensive and includes major bottlenecks.

Two examples show the extent of the challenge. For generic orbits in Kerr spacetime, calculating the first-order self-force at a single point in the parameter space can take $\sim 10^4$ CPU hours, requiring $\sim 10^4$ Fourier modes and $40+$ digits of precision~\cite{vandeMeent:2017bcc}. At second order, the problem grows: even in the simplest case of quasicircular, nonspinning binaries, calculating the inputs for the post-adiabatic waveform generation in Ref.~\cite{Wardell:2021fyy} required $\sim 10^6$ CPU hours.

These prohibitive expenses stem from a clash between two features of self-force calculations. First, to efficiently solve the field equations, we typically decompose the metric perturbations in a basis of angular harmonics defined on spheres centered on the large black hole and a basis of Fourier harmonics defined by the background's time-translation symmetry~\cite{Wardell15,Pound:2021qin}. Second, we treat the small companion as a point particle, which serves to eliminate the nonessential small-scale physics in and around the companion, but at the cost of introducing a singularity at the particle's position~\cite{Pound:2015tma,Pound:2021qin}. These two aspects, mode decompositions and singularities, are at odds with each other because singularities lead to slow convergence of mode sums. Even if one were to use a small, extended object rather than a point particle, the same difficulties would arise: very high modes would be required to accurately capture the steep field gradients in the object's vicinity (and even higher modes to capture the fields in its interior).

Our primary goal in this paper is to take a step toward reducing these obstacles. The method we propose is to radically smooth the fields in the problem. 

The idea underlying our proposal is a puncture scheme, which is a standard method in self-force calculations~\cite{Barack:2007we, Vega:2007mc, Barack:2009ux, Wardell15,  Barack:2018yvs, Pound:2021qin}. In a puncture scheme, one first constructs an analytical ``puncture field'' that captures the local singularity structure of the physical field in a neighborhood of the particle. One then subtracts this from the physical field to obtain a ``residual field''. The residual field satisfies a field equation with an ``effective source'' that is smoother than the original, physical source. We review the formulation of a puncture scheme in Sec.~\ref{sec:puncture_scheme}.

Puncture fields are derived as analytical expansions in powers of distance from the particle. The higher order the expansion is carried to, the smoother the residual field and effective source.  Our aim in this paper is to push puncture calculations to significantly higher order than has been done in the past. Since the numerical obstacles in existing calculations are linked to nonsmoothness, such higher-order punctures should reduce those obstacles. We explain the problems in more detail, and how a high-order puncture can help resolve them, in Sec.~\ref{sec:advantages}. There we also outline several other potential advantages of a high-order puncture.

Traditionally, punctures have been derived as approximations to the Detweiler-Whiting singular field, which is defined in terms of a particular Green's function~\cite{Detweiler03,Poisson:2011nh}. Obtaining a puncture then involves a variety of specialist technical tools: a Hadamard decomposition of the Green's function, followed by its near-coincidence expansion, leading to a local, covariant expansion of the singular field near the particle, and finally a re-expansion into a desired system of global coordinates defined on the black hole background spacetime. This sequence of steps underlies not just calculations of punctures, but also the bulk of all self-force calculations, as they rely on subtracting the Detweiler-Whiting singular field in order to extract physical effects of the self-force~\cite{Barack:2018yvs}. 

Our secondary goal in this paper is to demystify the Detweiler-Whiting singular field for nonexperts. Unlike at first order, punctures in second-order perturbation theory have been derived by directly solving the field equations in a local region around the particle~\cite{Pound:2012nt,Pound12,Gralla:2012db,Pound:2015tma}. In particular, Refs.~\cite{Pound:2009sm, Pound12, Pound:2014xva} provided a constructive procedure for obtaining the singular field by solving the field equations in a class of local, co-moving coordinate systems centred on the particle. The co-moving coordinates in that case were designed to remain valid in a generic background spacetime and for a generic particle trajectory. Here we show that by following the same constructive procedure in a coordinate system adapted to one's particular problem, one can calculate the singular field straightforwardly and efficiently at both first and second order in perturbation theory. The outcome is a simple method of deriving a high-order expansion of the singular field without any specialist expertise.

We explain and demonstrate the method in the simple test case of a scalar charge on a circular orbit in Schwarzschild spacetime. In that context, we calculate the puncture field to 14th order in distance, which is 10 orders higher than the previous maximum~\cite{Heffernan12}. (See Refs.~\cite{Ottewill:2009uj,Heffernan:2012vj,Heffernan:2022zgc} for other work on obtaining high-order expansions of the singular field using traditional methods.) As an additional demonstration, after calculating the puncture in local, co-moving coordinates, we show how it can be straightforwardly decomposed into azimuthal $m$ modes, for use in an $m$-mode puncture scheme such as the ones in Refs.~\cite{Barack:2007we,Dolan:2010mt,Dolan:2011dx,Thornburg:2016msc,Dolan:2012jg,Osburn:2022bby}. The computation of the high-order puncture and effective source are implemented on \textit{Mathematica} and are available upon request.
A companion paper~\cite{paperII} employs this high-order puncture in a new, multi-domain spectral puncture scheme.

The paper is organized as follows. Section~\ref{sec:scalar_field_schwarzschild} reviews the formulation of a puncture scheme, the traditional method of deriving a puncture from the Detweiler-Whiting singular field, and the problems that can be overcome by utilizing higher-order punctures. Section~\ref{sec:new_scheme} describes the new, constructive procedure, and Sec.~\ref{Sec:Results} demonstrates its results. We conclude in Sec.~\ref{sec:conclusion} with a discussion of future applications. Some details are relegated to appendices, and we restrict our attention to the simple scalar problem throughout. We use geometric units with $G=c=1$, Greek letters for 4D spacetime indices, and Latin letters for 3D spatial indices.

\section{Puncture schemes, the Detweiler-Whiting singular field, and problems of slow convergence}
\label{sec:scalar_field_schwarzschild}

Our toy problem consists of a particle carrying a scalar charge $q$ moving on a circular geodesic orbit around a Schwarzschild black hole. In that context, we begin by reviewing the basics of a puncture scheme and explain the potential utility of a high-order puncture. 


\subsection{Formulation of a puncture scheme}
\label{sec:puncture_scheme}

We consider a Schwarzschild spacetime with mass $M$ in standard
Schwarzschild coordinates $(t,r,\theta, \varphi)$.
The metric takes the familiar form
\begin{align}
\label{ss_metric}
\!\!ds^2 &= -f(r) dt^2 + f^{-1}(r) dr^2 + r^2 \bracket{d\theta^2 + \sin^2 \theta d\varphi^2}\!,
\end{align}
with $f(r) \coloneqq 1- \frac{2M}{r}$.

On this spacetime, we consider a particle whose worldline and four-velocity are given by $\xpart(\tau)$ and $u^\mu \coloneqq d \xpart/d\tau$ respectively, where $\tau$ denotes the particle's proper time.
We assume the particle moves on a circular geodesic at fixed radius $r=r_p$. Without loss of generality, we further restrict the motion to be in the equatorial plane, $\theta = \pi/2$.
$\xpart$ is obtained from the geodesic equation and is explicitly given by
\begin{equation}
\xpart = u^t \bracket{\tau,r_p,\frac{\pi}{2}, \Omega_p \tau},
\end{equation}
where
\begin{equation}
u^t = \frac{1}{\sqrt{1-\frac{3M}{r_p}}}, \quad \Omega_p \coloneqq \sqrt{\frac{M}{r_p^3}}.
\end{equation}
For the rest of this paper, quantities with a subscript $p$ imply their evaluation at the particle.

The charge sources a scalar field $\Phi$, which obeys the Klein-Gordon equation
\begin{equation}
\label{ScalarEqn}
\square \Phi  \coloneqq \nabla_\alpha \nabla^\alpha \Phi = - 4\pi \rho,
\end{equation}
where $\nabla$ is the covariant derivative with respect to the background metric \eqref{ss_metric} and $\rho$ is the particle's scalar charge density, given by
\begin{align}
\rho(t,r,\theta,\varphi) \coloneqq{}& q \int \frac{\delta^4(x^\mu-\xpart(\tau))}{\sqrt{-g}} d \tau \nonumber \\
    ={}& \frac{q}{r_p^2 u^t} \delta(r-r_p) \delta(\theta-\pi/2) \delta\bracket{\varphi-\Omega_p t}. \label{ScalarChargeDensity}
\end{align}


The puncture approach is based on a split of the full, retarded field $\Phiret$ into a singular, $\PhiS$, and regular, $\PhiR$, piece,
\begin{equation}
	\Phiret = \PhiS + \PhiR. 
\end{equation}
The singular piece is a particular solution of  Eq.~\eqref{ScalarEqn} and is singular at the particle. The regular field is instead a smooth solution of the homogeneous equation $\Box\PhiR=0$.
In general, this split is not unique since one can always add a homogeneous solution to the definition of $\PhiS$. However,  a judicious split~\cite{Detweiler03} makes it possible
to express the self-force only in terms of the regular field:
\begin{equation}
	F_\mu^{\rm self} = q \nabla_\mu \PhiR|_{x^\nu_p}. \label{SelfForceDef}
\end{equation}

In practice, it is generally not possible to obtain an exact, closed-form expression for $\PhiS$. What is possible is to obtain a local expression for $\PhiS$, written as an expansion in powers of distance to the particle. This local expansion, truncated at some finite order, is the puncture field, $\PhiP$.
The difference between the retarded and puncture field, $\Phires \coloneqq \Phiret-\PhiP$, is  the residual field. 
The residual field satisfies the Klein-Gordon equation with an effective source,
\begin{equation}
	\square \Phires = S^\text{eff} \coloneqq -4\pi \rho - \square \PhiP. \label{ResFieldEqnTMP}
\end{equation}
Starting from this basic idea, there are several ways to implement a puncture scheme. The most common method is to use $\Phires$ as the numerical variable inside a region $\Gamma$ around the particle's worldline and to use $\Phiret$ as the numerical variable outside $\Gamma$. One then solves Eq.~\eqref{ResFieldEqnTMP} inside $\Gamma$ and the homogeneous equation $\Box\Phiret=0$ outside $\Gamma$, subject to a junction condition on~$\Gamma$: $\Phiret=\Phires+\PhiP$ and  $\partial_n\Phiret=\partial_n\Phires+\partial_n\PhiP$, where $\partial_n$ denotes the derivative normal to $\Gamma$.  

The degrees of differentiability of $\Phires$ and $S^{\rm eff}$ depend directly on the order of the puncture~\cite{Dolan:2010mt}. If $\PhiP$ includes all terms in $\PhiS$ through $n$th order in distance, we refer to it as an $n$th-order puncture. When evaluated at the particle, the corresponding residual field $\Phires$ and its first $n$ derivatives agree with the regular field $\PhiR$ and its first $n$ derivatives. $\Phires$  is therefore a $C^n$ function at the particle's location, and the effective source $S^{\rm eff}$ is a $C^{n-2}$ function there.

\subsection{Local expansion of \texorpdfstring{$\PhiP$}{Φ{\textasciicircum}P} via the Detweiler-Whiting Green's function}\label{local expansion of PhiS}

To motivate our method of calculating $\PhiP$, we first review the traditional method of calculating it. The final output is a local expansion in powers of coordinate distance to the particle, as displayed in Eq.~\eqref{PhiS inertial coordinate expansion} or \eqref{PhiS inertial STF} below. Our goal here is only to outline the main ideas and mathematical tools, not the details, because one of our themes is that these ideas and tools are not essential for understanding and computing punctures. We refer the reader to \cite{Heffernan12,Poisson:2011nh} and references therein for more details.

The starting point is the Detweiler-Whiting Green's function for the operator $\Box$~\cite{Poisson:2011nh}: 
\begin{equation}
G^{\rm S}(x,x') := \frac{1}{2}\bigl[G^{ret}(x,x') + G^{adv}(x,x')-V(x,x')\bigr], 
\end{equation}
where $x$ is the field point and $x'$ is the source point. The retarded Green's function, $G^{ret}(x,x')$, is only nonzero if $x$ is in the causal future of $x'$, corresponding to causal, future-directed propagation of signals from the source; conversely, the advanced Green's function, $G^{adv}(x,x')$, is only nonzero if $x$ is within the causal past of $x'$, corresponding to past-directed propagation. The quantity $V(x,x')$ is a specific solution to the homogeneous equation $\Box V = 0$, chosen such that $G^{\rm S}$ is only nonzero if $x$ is \emph{outside} the chronological past and future of $x'$. If $x$ is within a convex normal neighbourhood of $x'$ (i.e., the region in which all points are connected to $x'$ by a unique geodesic), then $G^{\rm S}(x,x')$ can be written in the Hadamard form
\begin{equation}
	G^{\rm S}(x,x') = \frac{1}{2} \Bigl[ U(x,x') \delta(\sigma) - V(x,x') \Theta(\sigma) \Bigr].
\end{equation}
Here $\delta$ is the Dirac delta function, and $\Theta$ is the Heaviside function. The quantity $\sigma=\sigma(x,x')$ is Synge's world function~\cite{Synge1960}, defined to be half of the squared proper distance between $x'$ and $x$,
\begin{equation}
	\sigma(x,x') \coloneqq \frac{1}{2} \bracket{\int_\beta ds}^2,
\end{equation}
where $\beta$ is the unique geodesic connecting $x'$ to $x$, and $s$ is proper length along $\beta$. The key properties of $G^{\rm S}(x,x')$ are that it vanishes if $x$ and $x'$ are timelike separated and that it is symmetric in its arguments. The latter property follows from the symmetry of the two-point functions $U$ and $V$ in their arguments.

The singular field is the particular solution to Eq.~\eqref{ScalarEqn} defined by the Detweiler-Whiting Green's function:
\begin{align}
	\PhiS(x) &= \int G^{\rm S}(x,x')\rho(x')dV'\\
 &= \left.\frac{qU(x,x_p(\tau))}{2u^{\alpha'}\sigma_{\alpha'}}\right|^{\tau_{ret}}_{\tau_{adv}} -\frac{q}{2}\int^{\tau_{adv}}_{\tau_{ret}} V(x,x_p(\tau)) d\tau, \label{IntGreen}
\end{align}
where $x'=x_p(\tau)$ and we have introduced $\sigma_{\alpha'}:=\nabla_{\alpha'}\sigma$, which is a directed measure of proper distance from $x$ to $x'$. We will also use the natural generalisation $\sigma_{\alpha_1'\cdots\alpha_n'}:=\nabla_{\alpha_1'}\cdots\nabla_{\alpha_n'}\sigma$. We have additionally introduced the advanced time $\tau_{adv}$ as the time at the point $x'_{adv}=x_p(\tau_{adv})$ that is connected to $x$ by a past-directed null geodesic emanating from $x'_{adv}$; the retarded time is defined analogously in terms of a future-directed null geodesic emanating from $x'_{ret}=x_p(\tau_{ret})$. We note the first term in Eq.~\eqref{IntGreen} is derived from 
\begin{equation}
\int U(x,x_p(\tau))\delta(\sigma(x,x_p(\tau)))d\tau = \sum_{\substack{\tau=\tau_{ret},\\
\hfill\tau_{adv}}}\frac{U(x,x_p(\tau))}{|d\sigma/d\tau|}
\end{equation}
together with the fact that $d\sigma/d\tau>0$ at $\tau=\tau_{ret}$ and $d\sigma/d\tau<0$ at $\tau=\tau_{adv}$.

Traditionally, a local expansion of $\PhiS$ is obtained starting from Eq.~\eqref{IntGreen}. The two-point function $U(x,x')$ is given in terms of $\sigma$ and the bitensor of parallel transport, $g_\alpha^{\alpha'}(x,x')$, and it can be locally expanded in the coincidence limit $x\to x'$. Here a bitensor is an object that transforms as a tensor under coordinate transformations at $x$ and, separately, as a tensor at $x'$ under coordinate transformations there. The two-point function $V(x,x')$ is expressed as a Hadamard expansion,
\begin{equation}
	V(x,x') \coloneqq \sum_{n=0}^{\infty} V_n(x,x') \sigma^n(x,x').
\end{equation}
The coefficient $V_n$ are obtained from certain recursion relations, and all quantities are then expanded near coincidence.

Here we omit the technical tools involved in these expansions of $U$ and $V$ and instead focus on the generic features of the local expansion of $\PhiS$. The only properties we require for this are the facts that $U$ and $V$ are smooth functions of their arguments. We first choose a reference point $\bar x=x_p(\bar\tau)$ on the particle's worldline and define $\Delta\tau':=\tau' - \bar\tau$. We then expand functions at $x_p(\tau')$ around their values at $\bar\tau$. For example, 
\begin{align}
    V(x,x_p(\tau')) &= \sum_{k\geq0}\frac{1}{k!}(\Delta\tau')^k\frac{d^k}{d\bar\tau^k}V(x,x_0(\bar\tau)) \label{eq:V dtau expansion}
\end{align}
and similarly for $U(x,x_p(\tau'))$ and for $u^{\alpha'}\sigma_{\alpha'}$. Given the expansion~\eqref{eq:V dtau expansion}, the integral in Eq.~\eqref{IntGreen} evaluates to another series in $\Delta\tau'$,
\begin{multline}
\int^{\tau_{adv}}_{\tau_{ret}} V(x,x_p(\tau)) d\tau\\
=\sum_{k\geq0}\frac{(\Delta\tau_{adv})^{k+1}-(\Delta\tau_{ret})^{k+1}}{(k+1)!}\frac{d^k}{d\bar\tau^k}V(x,x_p(\bar\tau)).\label{int of V}
\end{multline}
These expansions are written in terms of covariant derivatives using $d/d\bar\tau = u^{\bar\alpha}\nabla_{\bar\alpha}$. Since $V(x,\bar x)$ and $U(x,\bar x)$ are smooth two-point functions, they have near-coincidence expansions of the form $\sum_{k\geq0}A^{\bar\alpha_1\cdots\bar\alpha_k}(\bar x)\sigma_{\bar\alpha_1}\cdots \sigma_{\bar\alpha_k}$ for some $A^{\bar\alpha_1\cdots\bar\alpha_k}$~\cite{Poisson:2011nh}.

Completing the expansions requires the quantities $\Delta\tau_{adv/ret}$, which can be found by expanding the relation
\begin{equation}
\sigma(x,x_p(\bar\tau+\Delta\tau_{adv/ret}))=0
\end{equation}
in powers of $\Delta\tau_{adv/ret}$. This leads to a series
\begin{align}\label{Dtau eqn}
    0 &= \sum_{k\geq0}\frac{1}{k!}(\Delta\tau)^k u^{\bar\alpha_1}\nabla_{\bar \alpha_1}\left[\cdots u^{\bar \alpha_k}\nabla_{\bar \alpha_k}\sigma(x,\bar x)\right]\\
    &= \sigma(x,\bar x) + \Delta\tau\, u^{\bar \alpha}\sigma_{\bar \alpha} +\frac{1}{2}(\Delta\tau)^2(a^{\bar \alpha}\sigma_{\bar \alpha} + u^{\bar \alpha}u^{\bar \beta}\sigma_{\bar \alpha\bar \beta})\nonumber\\
    &\qquad\qquad\qquad\qquad\quad\  +\ldots,
\end{align}
where $a^{\alpha}:=u^\beta\nabla u^\alpha$ is the covariant acceleration (which we set to zero in the bulk of this paper but leave nonzero here for generality). To solve Eq.~\eqref{Dtau eqn}, it is helpful to introduce a parameter $\lambda:=1$ that counts powers of distance from the particle. Second and higher derivatives of $\sigma$ can be re-expanded in a covariant Taylor series in the vector $\sigma_{\bar \alpha}$~\cite{Poisson:2011nh}. In particular,
\begin{equation}
    \sigma_{\bar\alpha\bar\beta} = g_{\bar\alpha\bar\beta} + \sum_{k\geq2}\lambda^k p^{\bar\alpha_1\cdots\bar\alpha_k}(\bar x)\sigma_{\bar\alpha_1}\cdots\sigma_{\bar\alpha_k}
\end{equation}
for some tensors $p^{\bar\alpha_1\cdots\bar\alpha_k}(\bar x)$. Expanding $\Delta\tau_{adv/ret}$ in powers of $\lambda$, as $\Delta\tau_{adv/ret} = \sum_{k\geq1}\lambda^k\Delta\tau^{(k)}_{adv/ret}$, and solving Eq.~\eqref{Dtau eqn} order by order in $\lambda$ then leads to 
\begin{equation}
    \Delta\tau^{(1)}_{adv/ret} = {\sf r} + \varepsilon{\sf s}
\end{equation}
where $\varepsilon=1$ for the advanced time and $-1$ for the retarded time.  The quantities
\begin{equation}
    {\sf r}:= u^{\bar \alpha}\sigma_{\bar \alpha}\quad\text{and}\quad {\sf s}:=\sqrt{(g^{\bar \alpha\bar \beta}+u^{\bar \alpha}u^{\bar \beta})\sigma_{\bar \alpha}\sigma_{\bar \beta}}
\end{equation}
are the components of $\sigma_{\bar\alpha}$ parallel to and orthogonal to the worldline, respectively, and we have used~\cite{Poisson:2011nh} 
\begin{equation}
2\sigma=\sigma_{\bar\alpha}\sigma^{\bar\alpha}={\sf s}^2 - {\sf r}^2.
\end{equation}
Subleading terms satisfy equations of the form
\begin{equation}
    \varepsilon s\Delta\tau^{(k)}_{adv/ret} = f_k(\Delta\tau^{(1)},\ldots,\Delta\tau^{(k-1)},\sigma_{\bar\alpha})
\end{equation}
for some $f_k$ that is a polynomial (of total order $\sim \lambda^{k+1}$) in the lower-order $\Delta\tau^{(k)}$'s and in contractions of $\sigma_{\bar\alpha}$ with available tensors (the Riemann tensor and its derivatives, the four-velocity, and the acceleration). We can observe, by induction, that $f_k$ contains terms with negative powers of $\varepsilon {\sf s}$ up to a maximum negative power $1/(\varepsilon {\sf s})^{k-2}$. Therefore
\begin{align}\label{dtau v1}
    \Delta\tau_{ret/adv} = {\sf r} + \varepsilon{\sf s} +\sum_{k\geq2}\lambda^k \frac{Q_k(\varepsilon\sf s,\sigma_{\bar\alpha})}{(\varepsilon{\sf s})^{k-1}},
\end{align}
for some $Q_k$ that is a polynomial (of total order $\sim \lambda^{2k-1}$) in $\varepsilon s$ and in contractions of $\sigma_{\bar\alpha}$ with available tensors.
We can write this in the alternative form
\begin{multline}
    \Delta\tau_{ret/adv} = {\sf r} + \varepsilon{\sf s} +\sum_{k\geq2}\lambda^k \biggl[\frac{q_1^{\bar\alpha_1\cdots\bar\alpha_{2k}}(\bar x)\sigma_{\bar\alpha_1}\cdots\sigma_{\bar\alpha_{2k}}}{(\varepsilon{\sf s})^{k}}\\
    +\frac{q_2^{\bar\alpha_1\cdots\bar\alpha_{2k-1}}(\bar x)\sigma_{\bar\alpha_1}\cdots\sigma_{\bar\alpha_{2k-1}}}{(\varepsilon{\sf s})^{k-1}}\biggr]\label{dtau v2}
\end{multline}
by noting that $Q_k$ is necessarily a sum of terms of the form $(\varepsilon {\sf s})^{j}t^{\bar\alpha_1\cdots\bar\alpha_{2k-1-j}}\sigma_{\bar\alpha_1}\cdots\sigma_{\bar\alpha_{2k-1-j}}$ for $0\leq j\leq 2k-1$. If $j$ is odd, we can multiply the numerator and denominator by $(\varepsilon{\sf s})$ to bring the numerator into the form of a smooth polynomial in $\sigma_{\bar\alpha}$; if $j$ is even, the numerator is already of that form. Obtaining a common denominator for all the even-$j$ terms and a common denominator for all the odd-$j$ terms then leads to the form~\eqref{dtau v2}.

Using the result~\eqref{dtau v1}  for $\Delta\tau_{ret/adv}$, we can now employ the strategy described around Eqs.~\eqref{eq:V dtau expansion} and~\eqref{int of V}: we expand functions of $(x,x')$ around $(x,\bar x)$, and we then expand them around the coincidence limit $x=\bar x$. We then arrive at expressions of the form
\begin{align}
    u^{\alpha'}\sigma_{\alpha'} &= {\sf r} + \Delta\tau_{adv/ret} (a^{\bar\alpha}\sigma_{\bar\alpha}+u^{\bar\alpha}u^{\bar\beta}\sigma_{\bar\alpha\bar\beta}) +{\cal O}(\lambda^2),\!\!\\
    &= -\varepsilon {\sf s}+\sum_{k\geq2}\lambda^k \frac{R_k(\varepsilon\sf s,\sigma_{\bar\alpha})}{(\varepsilon{\sf s})^{k-1}}
\end{align}
and
\begin{multline}
    U(x,x') = U(\bar x,\bar x) + ({\sf r}+\varepsilon{\sf s})\left[u^{\bar\alpha}\nabla_{\bar\alpha}U(x,\bar x)\right]_{x=\bar x} \\
    + \sum_{k\geq2}\lambda^k \frac{U_k(\varepsilon\sf s,\sigma_{\bar\alpha})}{(\varepsilon{\sf s})^{k-1}},
\end{multline}
where $R_k$ and $U_k$ are polynomials of the same form as $Q_k$, implying the sums can be rewritten in the form in Eq.~\eqref{dtau v2}. 
Together these results imply that the first term in the singular field~\eqref{IntGreen} has the local approximation
\begin{multline}
    \left.\frac{q U(x,x')}{2u^{\alpha'}\sigma_{\alpha'}}\right|^{\tau_{ret}}_{\tau_{adv}} \\ 
    = \frac{q}{{\sf s}} + \sum_{n\geq0}\lambda^n\,\frac{\Phi_1^{\bar \alpha_1\cdots\bar \alpha_{3n+3}}(\bar x)\sigma_{\bar \alpha_1}\cdots \sigma_{\bar \alpha_{3n+3}}}{{\sf s}^{2n+3}}.\label{first term expansion}
\end{multline}
Here we have used the explicit value $U(\bar x,\bar x)=1$~\cite{Poisson:2011nh} and the fact that even powers of $\varepsilon{\sf s}$ cancel between the retarded and advanced point. 

We obtain an analogous expansion of the second term in Eq.~\eqref{IntGreen} by combining \eqref{int of V} with Eq.~\eqref{dtau v2} and a near-coincidence expansion of $\frac{d^k}{d\bar\tau^k}V(x,x_p(\bar\tau))$. The result is of the form $\sum_{n\geq1}\lambda^n\,\frac{\Phi_2^{\bar \alpha_1\cdots\bar \alpha_{3n-3}}(\bar x)\sigma_{\bar \alpha_1}\cdots \sigma_{\bar \alpha_{3n-3}}}{{\sf s}^{2n-3}}$; multiplying numerator and denominator by ${\sf s}^6$ brings the result into the same form as~\eqref{first term expansion}.

Therefore, the singular field~\eqref{IntGreen} has a covariant expansion of the form
\begin{equation}\label{PhiS covariant expansion}
    \Phi^{\rm S} = \sum_{n\geq-1}\lambda^n\frac{P^{\bar \alpha_1\cdots\bar \alpha_{3n+3}}(\bar x)\sigma_{\bar \alpha_1}\cdots \sigma_{\bar \alpha_{3n+3}}}{{\sf s}^{2n+3}}.
\end{equation}
In fact, the terms with $n\geq0$ have ${\sf s}^{2n+1}$ in the denominator and $(3n+3)\to (3n+1)$ in the numerator, due to the explicit value $\nabla_{\bar\alpha}U(x,\bar x)|_{x=\bar x}=0$~\cite{Poisson:2011nh}. But to simplify the discussion below we can multiply both numerator and denominator by ${\sf s}^2$ to keep the generic form~\eqref{PhiS covariant expansion}. 

To express this in any given coordinate system, we reexpand for small coordinate distances $\Delta x^{\bar\alpha}:=x^\alpha - x^{\bar \alpha}$. At this stage, this only involves re-expressing $\sigma_{\bar\alpha}$ as an expansion in coordinate distance,
\begin{align}
    \sigma_{\bar \alpha} &= \sum_{n\geq1}s_{\bar \alpha \bar \beta_1\cdots \bar \beta_n}(\bar x)\Delta x^{\bar \beta_1}\cdots\Delta x^{\bar \beta_n} \\
    &= -g_{\bar\alpha\bar \beta}\Delta x^{\bar \beta} + \ldots 
\end{align}
We also introduce 
\begin{equation}
\rho:= \sqrt{(g_{\bar\alpha\bar\beta}+u_{\bar\alpha} u_{\bar \beta})\Delta x^{\bar\alpha}\Delta x^{\bar\beta}}
\end{equation}
as the leading term in the coordinate expansion of ${\sf s}$. Using 
\begin{equation}
    {\sf s} = \rho\sqrt{1+\sum_{k\geq 1}\lambda^k\frac{S_{\bar\alpha_1\cdots\bar\alpha_{k+2}}(\bar x)\Delta x^{\bar\alpha_1}\cdots\Delta x^{\bar\alpha_{k+2}}}{\rho^2}},
\end{equation}
we see that the covariant expansion~\eqref{PhiS covariant expansion} becomes
\begin{equation}\label{PhiS coordinate expansion}
       \Phi^{\rm S} = \sum_{n\geq-1}\frac{Q_{\bar\alpha_1\cdots\bar \alpha_{3n+3}}(\bar x)\Delta x^{\bar \alpha_1}\cdots \Delta x^{\bar \alpha_{3n+3}}}{\rho^{2n+3}}.
\end{equation}
Note that this same structure emerges simply from the expansion of the leading term, $\displaystyle\frac{q}{{\sf s}}$, in Eq.~\eqref{PhiS covariant expansion}; this motivated us to put the subleading terms in the form with a $1/{\sf s}^{2n+3}$ denominator, as described below Eq.~\eqref{PhiS covariant expansion}.

If we now specialise to a local coordinate system $(T,X^i)$, in which $X^i=0$ on the particle's worldline and $g_{\bar \alpha\bar \beta}={\rm diag}(-1,1,1,1)$ there, then $u^{\bar \alpha}=(1,0,0,0)$ and 
\begin{equation}
\rho = \sqrt{\delta_{ij}X^iX^j}:=R.    
\end{equation}
Introducing the unit vector $N^i := X^i/R$, we can then write Eq.~\eqref{PhiS coordinate expansion} as
\begin{equation}\label{PhiS inertial coordinate expansion}
       \Phi^{\rm S} = \sum_{n\geq-1}R^n Q_{i_1\cdots i_{3n+3}}(T)N^{i_1}\cdots N^{i_{3n+3}}.
\end{equation}
We make a final adjustment by re-writing the products of $N^i$'s in terms of symmetric trace-free (STF) combinations 
\begin{equation}
\hat N^L:=N^{\langle i_1}\cdots N^{i_\cl\rangle},    
\end{equation}
where traces are defined using $\delta_{ij}$. A product $N^{i_1}\cdots N^{i_\cl}$ can be written as a sum of terms involving $\hat N^{L}$, $\hat N^{L-2}$, $\hat N^{L-4}$, etc. For example, 
\begin{align}
N^i N^j &= \hat N^{ij}+\frac{1}{3}\delta^{ij},\\
N^i N^j N^k &= \hat N^{ijk}+\frac{1}{5}(\delta^{ij}N^k+\delta^{ik}N^j+\delta^{jk}N^i).
\end{align}
Hence, since $3n+3$ is odd for even $n$ and even for odd $n$, Eq.~\eqref{PhiS inertial coordinate expansion} can be rewritten as
\begin{equation}\label{PhiS inertial STF}
       \Phi^{\rm S} = \frac{q}{R}+\sum_{n\geq0}R^n \sum_{\substack{\cl \text{ even for odd }n\\\cl \text{ odd for even }n}}^{3n+3}\hat Q_L(T)\hat N^L
\end{equation}
for some STF tensors $\hat Q_L(T)$. We have pulled the leading term out of the sum to motivate expressions in later sections. Note that in those later sections, $\cl$ will be identified as the mode number in a spherical harmonic expansion defined on spheres centered on the particle; we reserve $\ell$ (with no bar) for the mode number in a spherical harmonic expansion defined on spheres centred on the black hole.

Equation~\eqref{PhiS inertial STF} is our key result in this section. Its critical feature is that it never features a term with $\cl=n$. As shown in Ref.~\cite{Pound12}, this feature suffices to uniquely specify its local form near the particle. By applying the method in Ref.~\cite{Pound12}, the next sections will make clear how the form in Eq.~\eqref{PhiS inertial STF} picks out a unique field. But we can also begin to understand it by contrasting Eq.~\eqref{PhiS inertial STF} with the form of the regular field. Since the regular field is smooth, it can be expanded in a Taylor series around the worldline,
\begin{equation}\label{PhiR inertial coordinate expansion}
       \Phi^{\rm R} = \sum_{n\geq0}R^n {\cal Q}_{i_1\cdots i_n}(T)N^{i_1}\cdots N^{i_n}.
\end{equation}
In terms of STF tensors,
\begin{equation}\label{PhiR inertial STF}
       \Phi^{\rm R} = \sum_{n\geq0}R^n \sum_{\substack{\cl \text{ even for even }n\\\cl \text{ odd for odd }n}}^n\hat{\cal Q}_L(T)\hat N^L.
\end{equation}
Each term $\propto R^n$ in this expansion has the opposite parity from the corresponding term in the singular field: under a parity transformation $X^i\to-X^i$, the terms of order $R^n$ are even parity for even $n$ and odd parity for odd $n$, while the terms of this order in the singular field have the opposite parity. And terms with $\cl=n$ always appear in the regular field, while they never appear in the singular field.

We also observe, tangentially, that the analysis in this section shows $\Phi^{\rm R}$, evaluated at the particle, is the Hadamard \emph{partie finie}~\cite{Blanchet:2000nu} of the full  field $\Phi=\Phi^{\rm S}+\Phi^{\rm R}$ on the particle. Similarly, the spatial derivatives of the regular field ($\partial_i\Phi^{\rm R}$, $\partial_i\partial_j\Phi^{\rm R}$, etc.), evaluated at the particle, are the \emph{partie finie} of the derivatives of the full field there. Here the Hadamard \emph{partie finie} of a singular function $\psi=\sum_{n}R^n \psi^{(n)}_L(T)\hat N^L$ at $R=0$ is defined to be $\frac{1}{4\pi}\oint\psi^{(0)}_{L}\hat N^L d\Omega = \psi^{(0)}_{0}$, the monopolar term in the coefficient of $R^0$. To the best of our knowledge, this connection between the Detweiler-Whiting regular field and Hadamard regularization has not been pointed out previously. We note, however, that Hadamard \emph{partie finie} regularization leads to ambiguous results at sufficiently high orders in post-Newtonian theory~\cite{Blanchet:2013haa}, and a naive application of it would likely lead to ambiguous results in second-order gravitational self-force theory due to the presence of $\log R$ terms in the retarded field~\cite{Pound:2012nt}; this will be explored in future work. 

\subsection{Accelerating convergence using high-order punctures}
\label{sec:advantages}

Before describing our new procedure, we briefly return to the relationship between the order of the puncture and the rate of convergence of numerical algorithms. 


Recall that for an $n$th-order puncture, $\Phires$ is  a $C^n$ function at the particle's location, and the effective source $S^{\rm eff}$ is a $C^{n-2}$ function there. If the fields are decomposed into modes, then typically each reduction in dimension translates into an increase in differentiability: if $\Phires$ is decomposed into $e^{im\varphi}$ modes, the coefficients $\Phires_{m}(t,r,\theta)$ are typically $C^{n+1}$  at the particle's position $(r_p,\pi/2)$; if $\Phires$ is decomposed into $Y_{\ell m}(\theta,\varphi)$ modes, the coefficients $\Phires_{\ell m}(t,r)$ are typically $C^{n+2}$ at $r_p$. The convergence rate of the mode coefficients is also directly linked to the order of the puncture. If $\Phires$ is $C^n$, then standard analysis~\cite{Orszag:1974,Miller:2020bft} tells us that, generically, its mode coefficients $\Phires_m$ decay at least as fast as $1/m^{n}$; likewise, its $\ell$-mode coefficients $\Phi_{\ell m}$ decay at least as fast as $1/\ell^{n-1}$.

In first-order self-force calculations, one is often only concerned with convergence properties far from the particle (to calculate gravitational-wave fluxes, for example) or precisely on the worldline (to calculate the self-force itself, for example). Mode convergence far from the particle is rapid. Calculations on the particle have most often used mode-sum regularization, in which one only requires the puncture's modes precisely at the particle, and the contribution of high-$\ell$ modes is easily approximated using a large-$\ell$ power-law fit for the high modes. In these situations, one is never interested in calculating the physical, retarded field in an entire region around the particle. Even so, there are potential advantages to using a puncture scheme with a high-order puncture, some of which we outline in the Conclusion. But there has been significantly less motivation than arises in second-order perturbation theory.

In second-order self-force calculations, the situation changes, primarily because there is a quadratic source term. This source, with the schematic form $\nabla h^{(1)}\nabla h^{(1)}$, is made up of products of the first-order retarded field $h^{(1)}_{\alpha\beta}$. Calculating this source from numerically computed modes of $h^{(1)}_{\alpha\beta}$ becomes impossible in a region of size $\sim M$ around the particle because the modes converge too slowly~\cite{Miller16}. To overcome this obstacle, second-order calculations~\cite{Pound:2019lzj,Warburton:2021kwk,Wardell:2021fyy} employ the strategy from Ref.~\cite{Miller16}: express the quadratic source as $\nabla h^{(1)\cal P}\nabla h^{(1)\cal P}+2\nabla h^{(1)\cal P}\nabla h^{(1)\cal R}+\nabla h^{(1)\cal R}\nabla h^{(1)\cal R}$, compute $\nabla h^{(1)\cal P}\nabla h^{(1)\cal P}$ from a 4D expression for $h^{(1)\cal P}_{\alpha\beta}$, and compute the remaining two terms using modes of $h^{(1)\cal P}_{\alpha\beta}$ and modes of $h^{(1)\cal R}_{\alpha\beta}$. This strategy requires exact integration of $h^{(1)\cal P}_{\alpha\beta}$ over a sphere to obtain its modes. At least currently, this integration can only be done numerically; the usual analytical methods of obtaining the $\ell m$ modes of $h^{(1)\cal P}_{\alpha\beta}$~\cite{Wardell:2015ada} involve additional approximations that cause the modes to diverge at large $\ell$ at all points away from the worldline~\cite{Toomani2022}. A high-order puncture could dramatically speed up this calculation by increasing the convergence rate of $\nabla h^{(1)\cal P}\nabla h^{(1)\cal R}$.

To understand the potential gain, consider that current second-order calculations required evaluating hundreds of millions of numerical integrals. This large number arises from the large number of modes (up to $\ell\sim 40$, multiplied by the number of $m$ modes and the 10 tensor components of the metric perturbation), orbital radii (ranging from $r_p=6M$ to $r_p\sim 25M$ at a spacing of $\sim 0.1M$), and radial grid points (several hundred in the range $r_p-2M<r<r_p+2M$ in which the puncture was used). A reduction in the number of required modes can therefore have a significant impact. Speeding up the convergence of the source calculation inside the worldtube could also allow one to increase the size of the worldtube, reducing the number of required modes of the retarded field outside the worldtube.

A high-order puncture for the second-order field, $h^{(2)\cal P}_{\alpha\beta}$, could compound this gain. By making the effective source smoother, a higher-order $h^{(2)\cal P}_{\alpha\beta}$ would reduce the need for a large number of grid points in the region around the particle. This last benefit has not been a major consideration at first order because the $\ell m$ modes of $h^{(1)\cal P}_{\alpha\beta}$ are finite at the particle and do not have steep gradients. At second order, the $\ell m$ modes of the physical source and of $h^{(2)\cal P}_{\alpha\beta}$ both diverge at $r=r_p$ and have steep gradients in a region around $r_p$.

We return to these issues and highlight other possible benefits of a high-order puncture in the Conclusion.

\section{New algorithm for computing the singular field}
\label{sec:new_scheme}

Traditionally, a puncture $\Phi^{\cal P}$ would be derived by starting from Eq.~\eqref{IntGreen} and carrying out the calculations sketched above to arrive at an expansion of the form~\eqref{PhiS coordinate expansion} in a convenient coordinate system. In this section we present a simple alternative: we directly calculate $\Phi^{\cal P}$ in a convenient coordinate system by substituting an ansatz of the form~\eqref{PhiS inertial STF} into the field equations, solving order by order in $R$, and then truncating at any desired $n_{\rm max}$.

Our method closely follows the approach presented by one of us in Ref.~\cite{Pound12}. However, that reference focused on local coordinate systems that can be constructed in any spacetime, such as Fermi-Walker coordinates. If starting from such a coordinate, one would then have to perform another expansion (or multiple expansions~\cite{Pound:2014xva}) to obtain the puncture in a coordinate system that is convenient for solving the field equations in the external spacetime (such as Schwarzschild coordinates centered on the large black hole, for example). Our new method instead calculates the puncture directly in a coordinate system that is straightforwardly related to the coordinates used in the external spacetime, without need for a re-expansion.


\subsection{Comoving coordinates}


Our first step is to construct useful comoving coordinates centered at the particle. Specifically, we will look for new coordinates $X^\mu \coloneqq \bracket{T,X,Y,Z}$ such that the following holds:
\begin{enumerate}
	\item The particle is located at $X=Y=Z=0$.
	\item The metric reduces to the flat metric $\eta_{\alpha \beta} = {\rm diag}(-1,1,1,1)$ at the particle.
\end{enumerate}
In what follows, the above are the only essential requirements for the coordinates $X^\mu$. Since the particle orbits around the black hole along a circular orbit at fixed radius, we will in addition demand for the coordinates to be comoving with the particle,
\begin{enumerate}
	\item[3.] $\bracket{\frac{\partial}{\partial T}}^\alpha = u^\alpha$ even away from the particle.
\end{enumerate}
This ensures that the scalar field (whether retarded, advanced, singular, or regular) is independent of the time coordinate $T$. 

These comoving coordinates can be found systematically by introducing a new set of basis vectors (tetrad) $e_\alpha \coloneqq e_\alpha^\mu \partial_\mu$, such that
\begin{equation}
g_{\alpha \beta}\big|_{\xpart} \coloneqq g(e_\alpha,e_\beta)\big|_{\xpart} \coloneqq \eta_{\alpha \beta}.
\end{equation}
A choice of tetrad is not unique, but is defined up to a Lorentz transformation; that is up to a $SO(1,3)$ transformation. This has six degrees of freedom. We demand one of the legs of the tetrad, say $e_1$, to be aligned with the four-velocity vector $u^\alpha$, leaving a residual freedom to rotate the legs $e_1$, $e_2$ and $e_3$ by an $SO(3)$ transformation. We will use the tetrad 
\begin{align}
    e_1 \coloneqq{}& u^\alpha = u^t \bracket{1,0,0,\Omega_p},\\
    e_2 \coloneqq{}& \bracket{0,\sqrt{f_p},0,0},\\
    e_3 \coloneqq{}& \bracket{0,0,\frac{1}{r_p},0},\\
    e_4 \coloneqq{}& -\frac{u^t}{r_p \sqrt{f_p}}\bracket{r_p^2 \Omega_p,0,0,f_p}.
\end{align}

\begin{figure}
    \centering
    \includegraphics[width=.95\columnwidth]{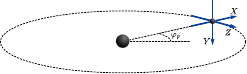}
    \caption{Local quasi-normal coordinates $X^i=(X,Y,Z)$ centered on the particle. In the black hole frame, the particle is on the equator ($\theta_p=\pi/2$) at a constant orbital radius $r_p$ and an azimuthal angle $\varphi_p=\Omega_p t$. Equation~\eqref{XYZ2rthetaphi} defines local spherical polar coordinates $(R,\ctheta,\cphi)$ from $X^i$ in the usual Euclidean way. Note that $Y$ is bounded between $-r_p$ (at $\theta=0$) and $+r_p$ (at $\theta=\pi$), and $Z$ is bounded between $z_c$ (at $\varphi=\varphi_p+\pi$) and $-z_c$ (at $\varphi=\varphi_p-\pi$), where $z_c<0$ is defined in Eq.~\eqref{zdef}. While $Z$ is not $2\pi$-periodic in $\varphi-\Omega_pt$, even functions of $Z$ (such as $\PhiP$) \emph{are} $2\pi$-periodic. The reader should also note that $X^i$ forms a left-handed coordinate system.}
    \label{fig:local-coords}
\end{figure}

We now want to define the comoving coordinates $X^\mu$ to be aligned with those basis vectors at the particle. That is, we impose
\begin{equation}
\left. \frac{\partial}{\partial X^\mu} \right|_\xpart \coloneqq e_\mu.
\end{equation}
Using the chain rule, this can be written as a system of PDEs,
\begin{equation}
\left. \frac{\partial x^\mu}{\partial X^\nu} \right|_\xpart = e_\nu^\mu, \qquad x^\mu(T,0,0,0)=\xpart.
\end{equation}
By construction, a solution will satisfy all three conditions given above.
Since the PDE system is only required to hold at the particle, there are uncountably many solutions. The solution we will use is given by
\begin{align}
    t \eqqcolon{}& e_1^1 T + e_4^1 z_c \sin^{-1} \bracket{\frac{Z}{z_c}},\\
    r \eqqcolon{}& r_p + e_2^2 X,\\
    \theta \eqqcolon{}& \cos^{-1} \bracket{-e_3^3 Y},\\
    \varphi \eqqcolon{}& e_1^4 T + e_4^4 z_c \sin^{-1} \bracket{\frac{Z}{z_c}},\\
    z_c \coloneqq{}& -2 r_p u^t \sqrt{f_p}.\label{zdef}
\end{align}
Inverting, we obtain the explicit relations
\begin{align}
	T &= u^t \bracket{f_p t - r_p^2 \Omega_p \varphi}, \label{T2BH}\\
	X &= \frac{r-r_p}{\sqrt{f_p}}, \label{X2BH} \\
	Y &= -r_p \cos \theta,\\
	Z &= z_c \sin \bracket{\frac{\varphi-\Omega_p t}{2}}. \label{Z2BH}
\end{align}
These local coordinates are illustrated in Fig.~\ref{fig:local-coords}.

In the above, the constant $z_c$ has been chosen so that, inverting the system, we have $Z \sim \sin \bracket{\frac{1}{2} (\varphi- \Omega_p t)}$. This factor of $1/2$ is important for the following reason. When considering spherical coordinates around the particle, the radius $R \coloneqq \sqrt{\delta_{ij}X^iX^j}$ should only vanish at the particle. Without the factor of $1/2$, there would instead be \textit{two} points at which $R=0$ in the Schwarzschild coordinates: the particle's position $(r_p,\pi/2,\varphi=\Omega_p t)$ and the antipodal point $(r_p,\pi/2,\varphi=\Omega_p t+\pi)$, leading to a spurious singularity at the antipode. This spurious singularity in turns severely complicates later calculations, which will involve integrals over the full range of $\varphi-\Omega_p t \in (-\pi,\pi]$. The factor of $1/2$ maps this spurious singularity back to $\varphi-\Omega_p t = 0$, where the first, physical one is located.

In those new coordinates, the non-trivial metric components read
\begin{subequations} \label{MetricComoving}
	\begin{align}
		g_{TT} &= -(u^t)^2 \bracket{f(r) - \Omega_p^2 \frac{r^2}{r_p^2} \bracket{r_p^2-Y^2}},\\
		g_{TZ} &= \frac{2 \Omega_p (u^t)^3}{\sqrt{z_c^2-Z^2}} \left[r_p^2 f(r) - f_p \frac{r^2}{r_p^2} (r_p^2-Y^2) \right],\\
		g_{XX} &= \frac{f_p}{f(r)},\\
		g_{YY} &= \frac{r^2}{r_p^2-Y^2},\\
		g_{ZZ} &= -\frac{4 (u^t)^4}{z_c^2-Z^2} \bracket{M r_p f(r) - f_p^2 \frac{r^2}{r_p^2} (r_p^2-Y^2)},
	\end{align}
\end{subequations}
where $r=r(X)$ is given by inverting \eqref{X2BH}.

\subsection{The field equation}

In the comoving coordinates defined above, the scalar charge density \eqref{ScalarChargeDensity} becomes
\begin{equation}
\rho = q \delta(X) \delta(Y) \delta(Z),
\end{equation}
as expected from local flatness.

Following the procedure given in Ref.~\cite{Pound12}, but adapted to the scalar field, we split the d'Alembert operator $\Box$ as follows:
\begin{align}
\Box \Phi &= g^{\alpha\beta} \nabla_\alpha \nabla_\beta \Phi,\\
&=g^{\alpha \beta} \bracket{\partial_\alpha \partial_\beta \Phi - \Gamma_{\alpha \beta}^\gamma \partial_\gamma \Phi},\\
&=\Delta_{\rm flat} \Phi - g^{\alpha \beta} \Gamma_{\alpha\beta}^\gamma \partial_\gamma \Phi  \nonumber \\
& \qquad +\bracket{g^{\alpha\beta}\partial_\alpha \partial_\beta -\Delta_{\rm flat}} \Phi,\\
& \eqqcolon \Delta_{\rm flat} \Phi - \cC(\Phi).
\end{align}
In the above, $\Delta_{\rm flat}=\partial^2_X+\partial^2_Y+\partial^2_Z$ is the 3D flat-space Laplacian.

$\cC$ can be re-written as
\begin{equation}
\label{Chat}
\cC = \hat{\delta} (g^{\alpha\beta} \Gamma_{\alpha\beta}^\gamma) \partial_\gamma - \hat{\delta}g^{\alpha\beta} \partial_\alpha \partial_\beta,
\end{equation}
where we defined $\hat\delta$ to denote the difference between a quantity and its flat-spacetime value; for example,
\begin{equation}
\hat{\delta}g^{\alpha \beta} \coloneqq g^{\alpha\beta}-\eta^{\alpha\beta},
\end{equation}
Note that for the comoving ``Cartesian'' coordinate $X^\mu$, we have that $\hat{\delta} \bracket{g^{\alpha \beta} \Gamma_{\alpha \beta}^\gamma} = g^{\alpha \beta} \Gamma_{\alpha\beta}^\gamma$.

We remark here that, interestingly, while the Christoffel symbols $\Gamma_{\alpha\beta}^\gamma$ in the coordinate $X^\mu$ are complicated, the particular combination $g^{\alpha \beta} \Gamma_{\alpha \beta}^c$ takes on a simple form, namely,
\begin{align}
	g^{\alpha \beta} \Gamma_{\alpha\beta}^\gamma &= \Gamma^\gamma_0 + \Gamma^\gamma_1 Z,\\
	\Gamma^\gamma_0 &\coloneqq \left\{0,-2\frac{r-M}{\sqrt{f_p} r^2}, \frac{2Y}{r^2},0 \right\},\\
	\Gamma^\gamma_1 &\coloneqq \left\{0,0,0, -\frac{M}{4 r_p^3 f} + \frac{r_p^2}{4 r^2 (r_p^2 - Y^2)} \right\}.
\end{align}
Here, again, $r=r(X)$ is defined from \eqref{X2BH}.

Returning now to \eqref{ScalarEqn}, the field equation reads
\begin{equation}
\Delta_{\rm flat} \Phi = -4\pi \rho + \cC(\Phi)
\end{equation}
We can cancel the delta function by extracting the corresponding singular behaviour in $\Phi$. Specifically, define
\begin{equation}
\Phi \eqqcolon \Phi^F + \frac{q}{\cR},
\end{equation}
where `F' is used to denote the part of the field that is finite at the particle. 
$\Phi^F$ satisfies the equation
\begin{equation}
\Delta_{\rm flat} \Phi^F = \cC\bracket{\Phi^F + \frac{q}{\cR}}.
\label{ETS}
\end{equation}

\subsection{Spectral decomposition and computation of the modes}
\label{section:Spectral_decomp_and_comput_modes}

The form~\eqref{PhiS inertial STF} is written in terms of STF combinations of unit vectors, where the unit vectors point outward from the particle. However, such a decomposition is in one-to-one correspondence with a spherical harmonic decomposition~\cite{Blanchet86}, with the $2\cl+1$ independent components of the STF $\hat N^L$ related to the $2\cl+1$ independent values of $\cm$ in $Y_{\cl \cm}$, where $Y_{\cl \cm}$ is defined in spheres of constant $R$ around the particle. We find it more convenient to work with the spherical harmonic representation.

We hence decompose $\Phi^F$ in terms of spherical coordinates around the particle,
\begin{equation}
\Phi^F = \sum_{n\geq0} \suml \summ \Phi_{\cl \cm \cn} \cR^\cn \Ylm(\ctheta,\cphi),
\label{phi_expansion}
\end{equation}
where
\begin{subequations} \label{XYZ2rthetaphi}
	\begin{align}
		X &= \cR \sin \ctheta \cos \cphi, \\
		Y &= \cR \sin \ctheta \sin \cphi, \\
		Z &= \cR \cos \ctheta,
	\end{align}
\end{subequations}
and where
\begin{equation}
\cl_\text{max} = 3 (\cn+1),\label{lmax}
\end{equation}
in accordance with Eq.~\eqref{PhiS inertial STF}. Note that we dropped the label ``F'' for the modes of~$\Phi^F$ since they are the same as the modes of the original field~$\Phi$.


The left-hand side of \eqref{ETS} then reads
\begin{multline}
\Delta_{\rm flat} \Phi^F = \sum_{n\geq0} \suml \summ \\
\left[\cn(\cn+1)-\cl(\cl+1)\right] \Phi_{\cl \cm \cn} \cR^{\cn-2} \Ylm.
\end{multline}
In the above, we made use of the well-known identity $\cR^2 \Delta_{\rm flat} \Ylm = -\cl(\cl+1) \Ylm$.

Similarly, by expanding the metric \eqref{MetricComoving} in powers of $\cR$, we can write the right-hand side of \eqref{ETS} as
\begin{equation}
\cC(\Phi) = \sum_{n\geq0} \suml \summ \cC_{\cl \cm \cn} \cR^{\cn-2} \Ylm(\ctheta,\cphi).
\label{CTerm}
\end{equation}
We can now solve $\Phi_{\cl \cm \cn}$ algebraically in terms of $\cC_{\cl \cm \cn}$ by equating coefficients. At any given value of $\cn$, there are two possibilities:
\begin{itemize}
	\item If $\cl \neq \cn$, then
	\begin{equation}
		\Phi_{\cl \cm \cn} = \frac{\cC_{\cl \cm \cn}}{\cn(\cn+1)-\cl(\cl+1)}. \label{PhilmnClmn}
	\end{equation}
	\item If $\cl=\cn$, then $\Phi_{\cn \cm \cn}$ is undetermined.
\end{itemize}
Note that the $\cC_{\cl \cm \cn}$ on the right-hand side of \eqref{PhilmnClmn} are themselves dependent on $\Phi_{\cl \cm \cn}$. However, by construction, they will only depend on $\Phi_{\cl \cm \cpp}$, where $\cpp < \cn$. This is because $\Delta_{\rm flat} \cR^\cpp \sim \cR^{\cpp-2}$, while $\cC(\cR^\cpp) \sim \cR^{\cpp-1} +$ higher-order terms. In particular, $\cC(\cR^\cpp)$ has no term $\sim \cR^{\cpp-2}$, since $\hat{\delta} g^{\alpha\beta} \sim \mathcal{O}(\cR)$. The resulting system of algebraic equations will therefore form a triangular system, where the modes at a given order $n$ are expressed as (linear) combinations of modes at previous orders. We show how to compute the $\cC_{\cl \cm \cn}$ in Appendix~\ref{sec:clmn}.

So, the solution is uniquely specified by the field equation at any order, except for the $\cl=\cn$ modes at each order $\cn\geq0$.
These ``homogeneous modes'' $\Phi_{\cn \cm \cn}$ contain all the freedom in the solution to the field equation.
As shown in Eq.~\eqref{PhiS inertial STF}, all of these homogeneous modes $\Phi_{\cn \cm \cn}$ are set to zero in the singular field. Therefore the singular field can be procedurally defined as follows:
\begin{align}
    \PhiS_{0,0,-1} &= 2 \sqrt{\pi} q,\label{PhiP00}\\
    \PhiS_{\cl \cm \cl} &=0, \label{PhiPnmn}\\
    \PhiS_{\cl \cm \cn} &= \frac{\cC_{\cl \cm \cn}}{\cn(\cn+1)-\cl(\cl+1)} \quad \text{for }\cl \neq \cn. \label{PhiPlmnClmn}
\end{align}
We emphasise that Eq.~\eqref{PhiPnmn} does not simply eliminate the homogeneous mode coefficients $\Phi_{\cn \cm \cn}$; it also eliminates all the pieces of the field that would otherwise, by virtue of Eq.~\eqref{PhiPlmnClmn}, be proportional to  $\Phi_{\cn \cm \cn}$. Each homogeneous mode acts as a ``seed'' that grows into an entire homogeneous solution via Eq.~\eqref{PhiPlmnClmn}~\cite{Pound:2015tma}. The singular field is the particular solution in which all of these seeded homogeneous solutions are set to zero. In the language of \cite{Pound12}, the singular field then corresponds to a \textit{minimal homogeneous solution} (which is homogeneous away from $R=0$) that is uniquely determined, via the algorithm \eqref{PhiP00}--\eqref{PhiPlmnClmn}, by its leading term, $q/R$.


\section{Results} \label{Sec:Results}

\subsection{The puncture field \texorpdfstring{$\PhiP$}{Φ{\textasciicircum}P}}

We implemented the algorithm~\eqref{PhiP00}--\eqref{PhiPlmnClmn} to compute $\PhiP$,
\begin{equation}
\label{eqn:PunctureFullDecomposition}
\PhiP = \sum_{\cn=-1}^{\cn_{\rm max}} \suml \summ \PhiP_{\cl \cm \cn} \cR^n \Ylm(\ctheta,\cphi),
\end{equation}
up to $\cn_{\rm max}=14$. This was achieved in a Mathematica notebook on a simple desktop computer (i5-7500 CPU \@ 3.4GHz). We have explicitly checked that the first five orders ($1/R$ through to $R^3$) of the resulting 4D puncture field $\PhiP$  agrees with  Ref.~\cite{Heffernan12}, where $\PhiP$ was obtained in covariant form up to and including order $\cR^4$. To make the comparison, we began with the covariant form~\eqref{PhiS covariant expansion} from Ref.~\cite{Heffernan12} and performed the local coordinate expansion as described in Sec.~\ref{local expansion of PhiS}. 



Since $\cl_{\rm max}$ obeys Eq.~\eqref{lmax}, computing all the non-vanishing modes at $\cn=14$ requires modes up to $\cl=45$. However, there are several simplifications that drastically reduce the actual number of modes that need to be computed. First, the structure~\eqref{PhiS inertial STF} dictates that 
\begin{equation}\label{l+n even = 0}
    \PhiP_{\cl\cm\cn} = 0\quad  \text{for even values of $\cl+\cn$}.
\end{equation}
As a result, only about half of the modes need to be computed.

Second, the singular field is symmetric under time reversal $(\varphi-\Omega_p t)\to - (\varphi-\Omega_p t)$~\cite{Hinderer:2008dm}, implying it must be an even function of $Z$. Since $Z=R\cos\ctheta$, this implies $\PhiP$ must be even under the reflection $\ctheta\to\pi-\ctheta$. Therefore,
\begin{equation}
    \PhiP_{\cl\cm\cn} = 0\quad  \text{for odd values of $\cl+\cm$}\label{l+m odd =0}
\end{equation}
because $\Ylm(\pi-\bar\theta,\bar\phi)=(-1)^{\cl+\cm}\Ylm(\bar\theta,\bar\phi)$. 

Next, since $\PhiP$ is real-valued and $Y_{\cl,-\cm}=(-1)^{\cm}(Y_{\cl\cm})^\star$, negative-$\cm$ modes can be calculated from positive-$\cm$ modes using
\begin{equation}\label{PhiP -m}
    \PhiP_{\cl,-\cm,\cn} = (-1)^{\cm}(\PhiP_{\cl\cm\cn})^\star,
\end{equation}
where the star denotes complex conjugation. We can  go further by noting that for a source orbit confined to the equator, the singular field possesses up-down symmetry (i.e., it is even under the reflection $\theta\to\pi-\theta$), which implies $\PhiP$ is an even function of $Y$. Since $Y=R\cos\ctheta\sin\cphi$, this in turn implies that $\PhiP$ must be an even function of $\cphi$ (at fixed $R$ and $\ctheta$). Given that the imaginary part of $\Ylm$ is an odd function of $\cphi$, we hence conclude that the coefficient of ${\rm Im}(\Ylm)$ must vanish. To determine that coefficient, we can rewrite the sum of $\cm\neq0$ terms as a sum over positive $\cm$ values,
\begin{equation}
    2\sum_{\cm=1}^\cl\left[{\rm Re}(\PhiP_{\cl\cm}){\rm Re}(\Ylm)-{\rm Im}(\PhiP_{\cl\cm}){\rm Im}(\Ylm)\right] ,
\end{equation}
where we have used Eq.~\eqref{PhiP -m} and the identity for $Y_{\cl,-\cm}$. Therefore
\begin{equation}
    {\rm Im}(\PhiP_{\cl\cm}) = 0,
\end{equation}
and Eq.~\eqref{PhiP -m} reduces to
\begin{equation}\label{PhiP -m real}
    \PhiP_{\cl,-\cm,\cn} = (-1)^{\cm}\PhiP_{\cl\cm\cn}.
\end{equation}

We note that our algorithm automatically enforces all of these conditions, but once they are known they can be used to streamline the computations. As an illustration, in Appendix~\ref{sec:PhilmnExpr} we give all the non-zero modes of $\PhiP_{\cl \cm \cn}$ for $\cn = -1,\cdots,2$.

\begin{figure} \centering
	\includegraphics[width=\columnwidth]{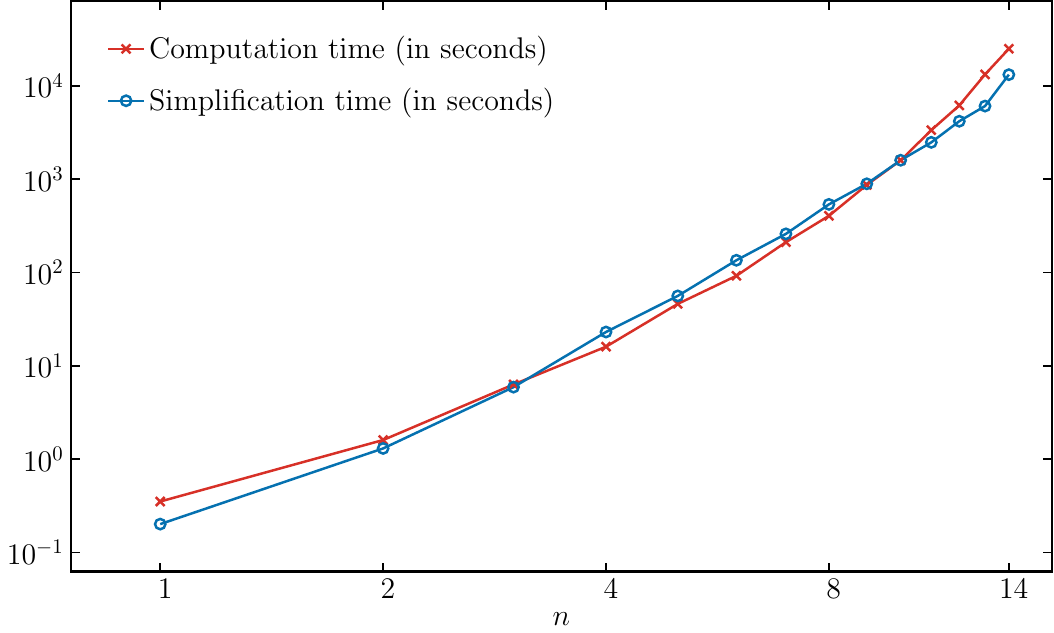}
	\caption{Log-log plot of the required  time for our algorithm to compute 	
		all the coefficients $\PhiP_{\cl \cm \cn}$ for each $\cn$, as well as their subsequent
		simplification.}
	\label{fig:compute_simplify_time}
\end{figure}

In Fig.~\ref{fig:compute_simplify_time} we plot the computation time to compute all coefficients $\PhiP_{\cl \cm \cn}$ at each $\cn$, given all the modes at lower $\cn$. We also plot the time required for Mathematica's built-in \textbf{Simplify} function to simplify the resulting expressions. 
After the first few orders, we can see that the computation and simplification time increase by approximately a factor of $2$ and $2.3$, respectively, from a given order to the next.

The reader may wonder about the overall ``size'' of the $\PhiP_{\cl \cm \cn}$ as one increases the puncture order. It turns out that the size of the expression  $\PhiP_{\cl \cm \cn}$ (say, the number of terms as functions of $M$ and $f_p$) increases linearly with $\cn$. Specifically, it goes approximately like $\sim 1.5 \cn$. The expressions therefore remain compact, and quick to evaluate, even for $\cn=14$. This surprising slow increase in the overall size of the modes can be attributed to the facts that (i) the number of modes also grows (quadratically) with $\cn$, and (ii) most of the general solution does not appear because the homogeneous modes $\Phi_{\cl \cm \cl}$ are set to zero. Regarding the second point, if the same algorithm is applied while keeping all the homogeneous modes arbitrary, we find that the size of the expression for each $\Phi_{\cl \cm\cn}$ grows as $\sim \cn^3$.


\subsection{\texorpdfstring{$m$}{m}-modes in the black-hole frame} \label{sec:mBH}


As alluded to in the Introduction, virtually all self-force calculations in black hole spacetimes make use of a mode decomposition~\cite{Barack:2009ux,Wardell15,Pound:2021qin} to partially or fully separate the field equations. The modes used in these calculations are defined on two-spheres centered on the black hole. It would therefore be useful to show how one can calculate the modes of our puncture field. In other words, one would like to be able to calculate the modes defined on the two-sphere around the black hole from the modes defined on the two-sphere around the particle. 


Calculating the $\ell m$ modes defined in the black hole's frame is highly nontrivial unless one makes additional approximations that spoil the convergence away from the particle; see Refs.~\cite{Miller16,Toomani2022,Bourg:2024vre}. We leave that mode decomposition to future work. However, it is reasonably straightforward to obtain exact analytical expressions for the $m$-modes of the puncture, at any order in $\cn$. These are the modes used in the $m$-mode scheme in the companion paper~\cite{paperII}.

For the special case of a circular orbit, an $m$-mode expansion is given by
\begin{equation}
    \PhiP(t,r,\theta,\varphi) = \sum_{m=-\infty}^\infty \PhiP_m(r,\theta)e^{im(\varphi-\Omega_p t)},
\end{equation}
where we have used the fact that the field can only depend on $\varphi$ and $t$ in the helically symmetric combination $\varphi-\Omega_p t$. The $m$-mode coefficients are given by 
\begin{equation}
\label{eqn:PhiPm}
\PhiP_m \coloneqq \frac{1}{2 \pi} \oint_{S^1} \PhiP e^{-im\varphi} d\varphi.
\end{equation}
In the integral, $t$ is set to zero or, equivalently, the integration variable $\varphi$ is to be interpreted as $\varphi-\Omega_p t$.

There are multiple ways to tackle the integral~\eqref{eqn:PhiPm} analytically.
In this paper, we will take an approach which will be most accommodating to evaluating $\PhiP_m$ (and its associated effective source) on a grid.

We start with the expression for the scalar field in spherical coordinates around the particle, $(\cR,\ctheta,\cphi)$:
\begin{equation}
	\PhiP = \sum_{\cl \cm \cn} \PhiP_{\cl\cm\cn} \cR^n Y_{\cl\cm}(\ctheta,\cphi).
\end{equation}
Plugging this decomposition into \eqref{eqn:PhiPm}, we have
\begin{equation}
	\PhiP_m =  \sum_{\cl\cm\cn} \frac{\PhiP_{\cl\cm\cn} c_{\cl\cm}}{2 \pi} \int_{-\pi}^{\pi} \cR^\cn P_\cl^\cm(\cos \ctheta) e^{i(\cm\cphi-m\varphi)} d \varphi,\label{PhimIntegral}
\end{equation}
where $c_{\cl\cm}$ is defined in Eq.~\eqref{clm}. 

It is useful to express all the quantities here in terms of the variables
\begin{align}
y &\coloneqq \cos^2 \theta,\\
\Delta r &\coloneqq r-r_p.
\end{align}
In particular, by equatorial symmetry, one expects all physical variables to be symmetric under the interchange $\theta \to \pi-\theta$. As a result, any dependence on $\cos \theta$ should appear squared, and we can restrict to the range $\cos \theta \geq 0$.

The spherical coordinates around the particle, $(\cR,\ctheta,\cphi)$, are easily related
to those in the black-hole frame $(t,r,\theta,\varphi)$ by using \eqref{XYZ2rthetaphi} and \eqref{T2BH}--\eqref{Z2BH}. 
We find
\begin{align}
	\cR^2 &= \varrho^2 + z_c^2 \sin^2 {\frac{\varphi}{2}}, \label{R to rho}\\
	\cos \ctheta &= \frac{Z}{\cR} = \frac{z_c \sin(\frac{\varphi}{2})}{\sqrt{\varrho^2+z_c^2 \sin^2(\frac{\varphi}{2})}}, \label{cos to rho}\\
	\tan \cphi &= \frac{Y}{X} = -\frac{r_p \sqrt{f_p y}}{r-r_p},
\end{align}
where we defined
\begin{equation}
	\varrho \coloneqq \sqrt{\frac{\Delta r^2}{f_p} + r_p^2 y}.
\end{equation}

Note that $\cphi$ is independent of $\varphi$. As a result, we can write \eqref{PhimIntegral} as
\begin{equation}
	\PhiP_m = \sum_{\cl\cm\cn} \frac{\PhiP_{\cl\cm\cn} c_{\cl\cm}}{2 \pi} e^{i\cm\cphi}\, \mathfrak{I}^m_{\cl\cm\cn}(\varrho).
\end{equation}
Here we defined
\begin{equation}\label{Imlmn def}
	\Ilmnm(\varrho) \coloneqq \int_{-\pi}^{\pi} \cR^\cn P_\cl^\cm(\cos \ctheta) e^{-im\varphi} d \varphi,
\end{equation}
where $\mathfrak{I}$ only depends on the Schwarzschild coordinates $(r,\theta)$ via $\varrho$.

We can further simplify the above expression by explicitly enforcing the reality condition on the scalar field, $\PhiP = (\PhiP)^\star$, which requires that $\PhiP_{\cl(-\cm)\cn} = (-1)^\cm \PhiP_{\cl\cm\cn}$; recall Eq.~\eqref{PhiP -m real}. This allows us to write
\begin{equation}\label{PhiP m formula}
	\PhiP_m = \sum_{\cn=-1}^{\cn_\text{max}} \sum_{\cl=0}^{3\cn+3} \sum_{\cm=0}^\cl \frac{\PhiP_{\cl\cm\cn} c_{\cl\cm}}{\pi (1+\delta_0^m)} \cos(\cm\cphi) \mathfrak{I}^m_{\cl\cm\cn}(\varrho).
\end{equation}
Finally, we can note that Eq.~\eqref{l+m odd =0} dictates that we only require $\mathfrak{I}_{\cl \cm \cn}^m$ for even values of $\cl+\cm$.

\begin{figure}
    \centering
    \includegraphics[width=0.8\columnwidth]{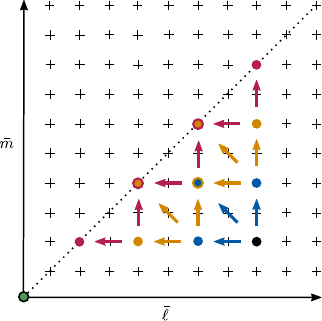}
    \caption{Structure of the recursion relation~\eqref{recursion_relation} and relationship~\eqref{top case 1}. Using ~\eqref{recursion_relation}, we express $\Ilmnm$ at the black point $(\cl,\cm)$ on the bottom right in terms of the integrals at the three blue points; the integrals at the blue points in terms of the integrals at the neighboring gold points; and the integrals at the gold points in terms of the integrals at the rose points. In this way, $\Ilmnm$ at the black point $(\cl,\cm)$ is expressed in terms of integrals at points on the dotted line $\cm=\cl$. Integrals at points on the dotted line are then expressed in terms of integrals at the origin $(0,0)$ (green point) using~\eqref{top case 1}.}
    \label{fig:triangle}
\end{figure}

There are several ways to find an analytic expression for $\mathfrak{I}$. The most efficient method we found is to make use of the following recursion relation for the associated Legendre polynomials:
\begin{multline}
	(\cl-\cm-1)(\cl-\cm)P_\cl^\cm(x) = P_{\cl-2}^{\cm+2}(x) \\
	   - P_\cl^{\cm+2}(x) + (\cl+\cm)(\cl+\cm-1)P_{\cl-2}^\cm(x).
\end{multline}
By linearity, $\mathfrak{I}$ then satisfies a similar relation,
\begin{multline}
	(\cl-\cm-1)(\cl-\cm) \Ilmnm = \mathfrak{I}_{(\cl-2) (\cm+2) \cn}^m \\
	   - \mathfrak{I}_{\cl (\cm+2) \cn}^m + (\cl+\cm)(\cl+\cm-1) \mathfrak{I}_{(\cl-2) \cm \cn}^m,
	\label{recursion_relation}
\end{multline}
such that $\mathfrak{I}^m_{\cl\cm\cn}$ can be expressed in terms of three other integrals at lower $\cl$ and/or higher $\cm$. This can be applied repeatedly to express $\mathfrak{I}^m_{\cl\cm\cn}$ in terms of $\mathfrak{I}^m_{\ell'm'\cn}$ for $(\ell',m')$ within the triangle $m'\leq\ell'\leq\cl$, $\cm\leq m'\leq\cl$, as illustrated in Fig.~\ref{fig:triangle}. The three-point recursion terminates at the ``top'' case $m'=\ell'$ since $\mathfrak{I}_{\ell'm' \cn}^m=0$ for $m'>\ell'$; the latter follows from the definition~\eqref{Imlmn def} and the fact that $P^{m'}_{\ell'}=0$ for integers $m'>\ell'$. Note that the three-point recursion~\eqref{recursion_relation}  always arrives at $m'=\ell'$, rather than stopping short at $m'=\ell'-1$, because we only require $\mathfrak{I}^m_{\cl\cm n}$ for even values of $\cl+\cm$, which in turn implies that $\ell'+m'$ is always even.

In order to close the recursion, we hence need to provide explicit formulas for the  case $\cl=\cm$. For that top case, we make use of the relation
\begin{align}
	P_\cl^\cl(x) &= (-1)^\cl (2\cl-1)!! (1-x^2)^{\cl/2}.
\end{align}
Substituting this into Eq.~\eqref{Imlmn def} and using Eqs.~\eqref{R to rho} and~\eqref{cos to rho}, we find
\begin{align}
	\mathfrak{I}_{\cl \cl \cn}^m &= (-1)^\cl (2\cl-1)!! \varrho^\cl \mathfrak{I}_{00 (\cn-\cl)}^m.\label{top case 1}
\end{align}
We therefore only require an explicit formula for the special case $\cl=\cm=0$.

For that special case, noting that the imaginary part vanishes by oddness of the integrand, we write
\begin{equation}
	\mathfrak{I}_{00 \cn}^m = \int_{-\pi}^{\pi} \bracket{\varrho^2 + z_c^2 \sin^2\bracket{\frac{\varphi}{2}}}^{\frac{\cn}{2}} \cos(m \varphi) d\varphi. \label{I00nmIntegral}
\end{equation}
Evaluated with Mathematica, this integral admits an explicit representation in terms of a regularized generalised hypergeometric function,
\begin{align}
	\mathfrak{I}_{00 \cn}^m &= 2 \pi (z_c^2 + \varrho^2)^\frac{\cn}{2} \times \nonumber \\
	&_3F_2^{\text{reg}}\bracket{\left\{\frac{1}{2},1,-\frac{\cn}{2}\right\},\{1-m,1+m\},\frac{z_c^2}{z_c^2+\varrho^2}}. \label{I003F2}
\end{align}
While the above function is well defined for all $\varrho > 0$, and even for non-integer values of $m$ and $\cn$, the evaluation time near $\varrho \simeq 0$ for certain parameter values $m$ and $\cn$ is particularly slow. In our case, however, we can make use of the fact that $m$ only admits integer values to find a much simpler analytic formula.
Specifically, the integral above can be recast into a form given in Eq.~(3.664.3) of \citet{Grad14}. Namely, for any integer $n$, and real number $q$, we have
\begin{equation}
	\int_{-\pi}^\pi \bracket{z+\sqrt{z^2-1} \cos \varphi}^q \cos (n\varphi) d\varphi = \frac{2 \pi}{(1+q)_n} {}_3 P_q^n(z), \label{eq:GR36643}
\end{equation}
where $(a)_n$ is the Pochhammer symbol.
In the above, $\phantom{}_3P_l^m(x)$ denotes the associated Legendre \emph{function}, defined for \(x>1\). We use the prescript ``\(3\)'' to match with Mathematica's terminology of referring to this as a ``type \(3\)'' associated Legendre function. This is in contrast to the associated Legendre \emph{polynomial}, \(P_l^m(x)\), defined for \(-1<x<1\).


The specific transformation to bring \eqref{I00nmIntegral} into the above form is
\begin{equation}
	z = \frac{z_c^2/2+\varrho^2}{\varrho \sqrt{z_c^2+\varrho^2}}.
\end{equation}
We then find that
\begin{equation}
	\mathfrak{I}_{00 \cn}^m \propto \int_{-\pi}^\pi \bracket{z-\sqrt{z^2-1} \cos \varphi}^\frac{\cn}{2} \cos (m\varphi) d\varphi. \label{eq:I00nIntegral}
\end{equation}
Making the substitution \(\varphi'=\varphi+\pi\), and noting that the integrand is an even function of \(\varphi'\) with period \(2\pi\), we can write Eq.~\eqref{eq:I00nIntegral} as
\begin{equation}
    \mathfrak{I}_{00 \cn}^m \propto 2(-1)^m\int_0^\pi\left(z+\sqrt{z^2-1}\cos\varphi'\right)^{\frac{n}{2}}\cos(m\varphi')d\varphi',
\end{equation}
which is the same form as Eq.~\eqref{eq:GR36643}.
This then gives the final result,
\begin{multline}
    \mathfrak{I}_{00 \cn}^m = \frac{2 \pi (-1)^m \bracket{\varrho \sqrt{z_c^2+\varrho^2}}^{\cn/2}}{(1+\cn/2)_m} \\
	   \times {}_3P_{\cn/2}^m \bracket{\frac{z_c^2/2+\varrho^2}{\varrho \sqrt{z_c^2+\varrho^2}}}. \label{I00nm}
\end{multline}

In summary, the $m$ modes of the puncture are given by Eq.~\eqref{PhiP m formula} with the recursion formulas \eqref{recursion_relation} and \eqref{top case 1}, starting from the extreme case~\eqref{I00nm}. The workflow to compute $\mathfrak{I}_{\cl \cm \cn}^m$ can therefore be summarized as follows:
\begin{enumerate}
    \item Apply Eq.~\eqref{recursion_relation} repeatedly until $\mathfrak{I}_{\cl \cm \cn}^m$ is expressed as a linear combination of $\mathfrak{I}_{\ell'\ell' \cn}^m$ for $\cm\leq\ell'\leq \cl$.
    \item Use Eq.~\eqref{top case 1} to reduce the $\mathfrak{I}_{\ell'\ell'\cn}^m$ terms to $\mathfrak{I}_{00 n'}^m$ for $(n-\cl)\leq n'\leq (n-\cm)$.
    \item 
    Use Eq.~\eqref{I00nm} to compute $\mathfrak{I}_{00n'}^m$.
\end{enumerate}

We can examine the singularity structure of the $m$ modes~\eqref{PhiP m formula} by adopting $(\varrho,\cphi)$ as polar coordinates centred on the particle in the $(r,\theta)$ plane. A function of the form~\eqref{PhiP m formula} is smooth at $\varrho=0$ if and only if $\mathfrak{I}_{\cl\cm\cn}^m(\varrho)\propto \varrho^{\cm+2k}$ for integer $k\geq0$. Equations~\eqref{top case 1} and~\eqref{I00nm}, together with the asymptotic approximations for the Legendre function~\cite{LegendreTypeIII}, imply the following behavior near $\varrho=0$: 
\begin{equation}
\mathfrak{I}_{\cl\cl\cn}^m(\varrho) = \begin{cases}
   {\cal O}(\varrho^{\cl}) &\text{for $n-\cl$ even},\\
    {\cal O}(\varrho^{\cl}) + {\cal O}(\varrho^{n+1}\log\varrho) &\text{for $n-\cl$ odd}.
\end{cases}
\end{equation}
The recursion relation~\eqref{recursion_relation}, combined with the fact that $\ell'\geq\cm$, then suggests that $\mathfrak{I}_{\cl\cm\cn}^m$ has the form
\begin{equation}
  \mathfrak{I}_{\cl\cm\cn}^m = \begin{cases}
   {\cal O}(\varrho^{\cm+2k}) &\text{for $n-\cl$ even},\\
    {\cal O}(\varrho^{\cm+2k}) + {\cal O}(\varrho^{n+1}\log\varrho) &\text{for $n-\cl$ odd}.
\end{cases}
\end{equation}
Here the ``$+2k$'' denotes the fact that subleading terms come as additional even powers of $\varrho$; this is because (i) subleading terms in the small-$\varrho$ expansion of Eq.~\eqref{I00nm} appear with even powers of $\varrho$ relative to the leading term, and (ii) the recursion relation~\eqref{recursion_relation} only moves in increments of two.

Recalling that $\varrho^{\cm+2k}\cos(\cm \cphi)$ is smooth, we can now see that the $m$ modes of the puncture, as given in Eq.~\eqref{PhiP m formula}, are made up of smooth terms plus terms of the form 
\begin{equation}\label{PhiPm singularity structure}
    \cos(\cm\cphi)\varrho^{n+1}\log\varrho
\end{equation}
near $\varrho=0$, with $\cn\geq -1$ and $0\leq\cm\leq 3n+3$. Also recalling Eqs.~\eqref{l+n even = 0} and~\eqref{l+m odd =0}, we observe that the puncture is only nonzero for $\cn-\cl=$~odd and $\cl+\cm=$~even, meaning the singular terms~\eqref{PhiPm singularity structure} only appear for $\cn+\cm=$~odd.

\subsection{The effective source \texorpdfstring{$S^\text{eff}$}{S{\textasciicircum}eff}}



The full, 4D puncture can be written as in \eqref{eqn:PunctureFullDecomposition}.
The full 4D effective source  is then obtained by simply applying the operator $-\square$ to $\PhiP$; the distributional content at the particle cancel out. Rather than belaboring the form of this 4D effective source, here we focus on the construction of the $m$-mode effective source, as used in the companion paper~\cite{paperII}. 

In order to compute the effective source's black-hole centered $m$ modes, we could evaluate
\begin{equation}
\label{eq:SeffmFull}
	S^{\rm eff}_m \coloneqq \frac{1}{2 \pi} \oint_{S^1} S^{\rm eff} e^{-i m\varphi} d\varphi = -\frac{1}{2 \pi} \oint_{S^1} e^{-i m\varphi} \square \PhiP d\varphi.
\end{equation}
However, we can more easily express $S^{\rm eff}_m$ directly in terms of the $m$-modes of $\PhiP$ by noting that, for a generic function $H(t,r,\theta,\varphi) = h(r,\theta) e^{-i m(\varphi - \Omega_p t)}$, we have
\begin{align}
\square H &= \nabla_\alpha \nabla^\alpha H = \frac{1}{\sqrt{-g}} \partial_\alpha \bracket{\sqrt{-g} g^{\alpha\beta} \partial_\beta H},\\
		&= (\square_m h) e^{-i m(\varphi - \Omega_p t)},
\end{align}
where $\square_m = {}^0\square_m + {}^1\square + {}^2\square$ and
\begin{align}
	{}^0\square_m &\coloneqq m^2 \bracket{\frac{\Omega_p^2}{f} - \frac{1}{r^2 (1-y)}}, \label{eqn:square0} \\
	{}^1\square &\coloneqq \frac{1+f}{r} \frac{\partial}{\partial r} + \frac{2 (1-3y)}{r^2} \frac{\partial}{\partial y}, \\
	{}^2\square &\coloneqq f \frac{\partial^2}{\partial r^2} + \frac{4y(1-y)}{r^2} \frac{\partial^2}{\partial y^2}. \label{eqn:square2}
\end{align}
In particular,
\begin{equation}
\bracket{\square H}_m = \square_m h.
\end{equation}
It then follows that \eqref{eq:SeffmFull} can be re-written as
\begin{equation}
S^{\rm eff}_m = -\square_m \PhiP_m.
\end{equation}

In Appendix~\ref{app:Seffm}, we give a detailed description of how to evaluate these formulas to compute $S^{\rm eff}_m$ analytically from $\PhiP_m$.

\begin{figure*} 
	\includegraphics[width=0.725\textwidth]{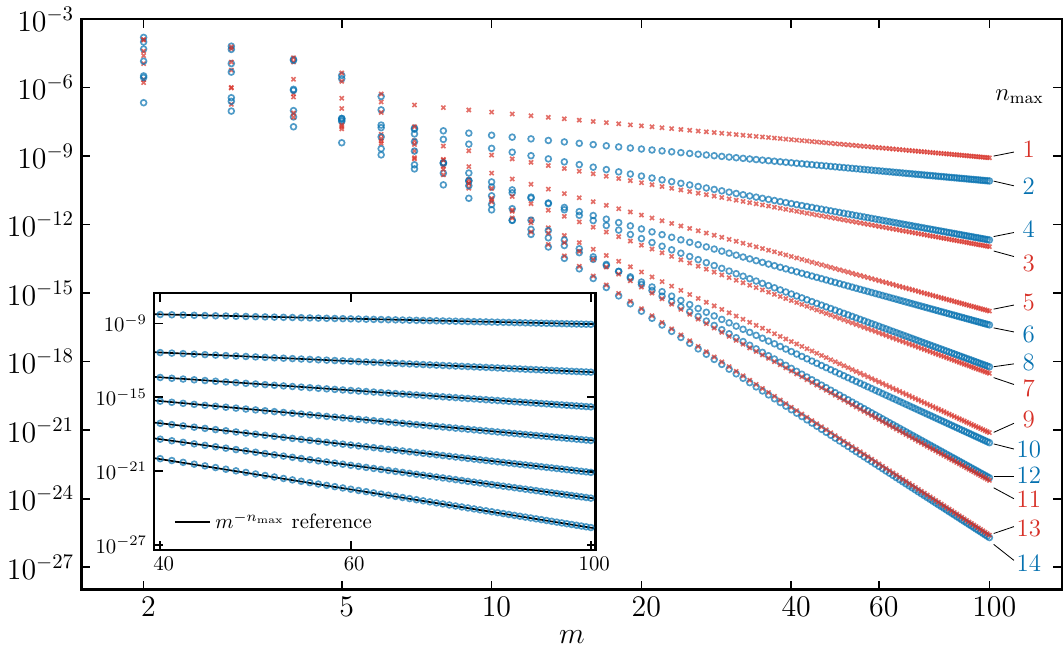}
	\caption{$m$-modes of the effective source, $|S^{\rm eff}_m|$,  for $\cn_\text{max}=1,\ldots,14$, evaluated near the particle, at $(r-r_p)/M = 10^{-6}=\cos^2\theta$.
	For large $m$, we recover the expected theoretical result. Namely, for even $\cn_\text{max}$, $|S^{\rm eff}_m| \sim m^{-\cn_\text{max}}$,
	while for odd $\cn_\text{max}$, $|S^{\rm eff}_m| \sim m^{-\cn_\text{max}-1}$. This is shown in the inset, where, for clarity,
	we only plot the $m$-mode contribution for even $\cn_\text{max}$.}
	\label{fig:Seff_mBehaviour}
\end{figure*}

Our primary purpose in going to large values of $\cn_\text{max}$ is to achieve more rapid numerical convergence. To assess the convergence with $m$, in Fig.~\ref{fig:Seff_mBehaviour} we show the asymptotic large-$m$ behavior of the effective source modes $S^{\rm eff}_m$ for different puncture orders. For these plots we evaluate $S^{\rm eff}_m$ near the particle, $(r-r_p)/M = 10^{-6} = \cos^2 \theta$. We avoid evaluating precisely at the particle because the formula for $S^{\rm eff}_m$, when evaluated at the particle, involves cancellations of quantities that diverge there. We remark that it may be possible to obtain a separate formula specifically tailored to compute the effective source at the particle, by analytically canceling out these divergences. 

Due to the form of the puncture, the asymptotic behaviour of the effective source (and puncture) near the particle has a polynomial behaviour with $m$. In fact, one can show that the convergence with $m$ improves by exactly two powers every second order in $\cn$~\cite{Barack:2007we}, as we can see in Fig.~\ref{fig:Seff_mBehaviour}. Specifically, starting at $\cn_\text{max}=1$, $S^{\rm eff}_m \sim m^{-\cn_\text{max}}$ for even $\cn_\text{max}$, while $S^{\rm eff}_m \sim m^{-\cn_\text{max}-1}$ for odd $\cn_\text{max}$.

As expected, we achieve very rapid convergence for large values of $\cn_\text{max}$: a 4th-order puncture (the highest previously available) requires 100 $m$-modes to achieve the same accuracy that a 14th-order puncture attains with only $\sim 15$ $m$-modes.

\subsection{Regularity of \texorpdfstring{$S^\text{eff}_m$}{S{\textasciicircum}eff{\_}m} at the particle}
\label{section:Regularity_Of_Seff}

In general, the regularity of the effective source at the particle is directly dependent on the order of the puncture~\cite{Dolan:2010mt}. 

In 4D, before performing the $m$-mode decomposition, the form of the effective source can be easily deduced from the form of the singular field in Eq.~\eqref{PhiS coordinate expansion}. Suppose we truncate the puncture at $\cn=\cn_{\rm max}$. The form of the effective source then corresponds to the d'Alembertian applied to the first omitted term, $\cn=\cn_{\rm max}+1$, in the singular field~\eqref{PhiS coordinate expansion}. We immediately find
\begin{equation}\label{Seff coordinate expansion}
       S^{\rm eff} = \sum_{n\geq\cn_{\rm max}+1}\frac{Q_{\bar\alpha_1\cdots\bar \alpha_{3n+3}}(\bar x)\Delta x^{\bar \alpha_1}\cdots \Delta x^{\bar \alpha_{3n+3}}}{\rho^{2n+5}}.
\end{equation}
This can be written in local polar coordinates $(\cR,\ctheta,\cphi)$ and then decomposed into $m$ modes in the same way as the puncture. In analogy with Eq.~\eqref{PhiPm singularity structure}, the dominant singularity in the $m$ modes $S^{\rm eff}_{m}$ is composed of terms of the form 
\begin{equation}\label{Seffm singularity structure}
    \cos(\cm\cphi)\varrho^{n_{\rm max}}\log\varrho \quad\text{with $\cn_{\rm max}+\cm=$~even}.
\end{equation}
The relationship between $\cm$ and $\cn_{\rm max}$ is a consequence of the fact that these terms come from analogous terms with $\cn=\cn_{\rm max}+1$ in Eq.~\eqref{PhiPm singularity structure}, which all satisfy $\cn+\cm=$~odd.

Equation~\eqref{Seffm singularity structure} shows that $S^{\rm eff}_m$ is $C^{\cn_{\rm max}-1}$ at the particle. Due to the appearance of $\cos(\cm\cphi)$ and the condition $\cn_{\rm max}+\cm=$~even, the singularity is reduced by one order for $\cn_{\rm max}=$~odd when approaching the particle along lines of constant $\cphi=\pm \pi/2$, orthogonal to the orbital plane.

\begin{figure*}{\centering
\includegraphics[width=\textwidth]{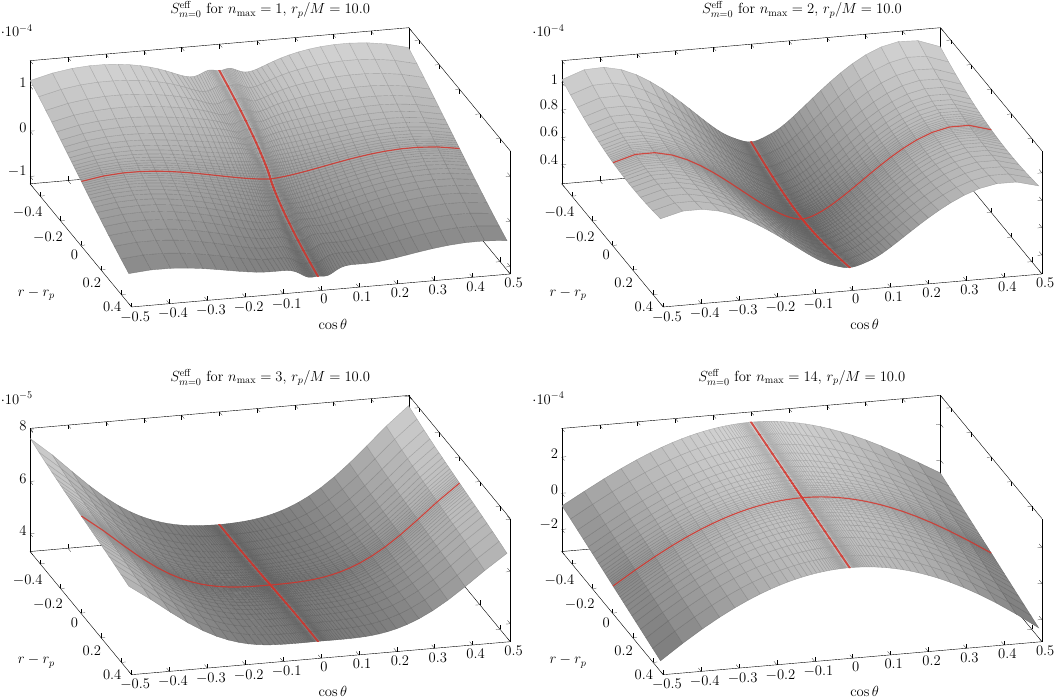}
%
 }
	\caption{The effective source $S^{\rm eff}_m$ for different puncture orders $\cn_{\rm max}$. In all cases, we restrict to $m=0$ and $r_p/M = 10.0$. For clarity, drawn in red are the lines $r=r_p$ and $\theta=\pi/2$. For $\cn_{\rm max}=1$ (top left) the effective source is continuous at the particle but not differentiable. Regularity at the particle improves as $\cn_{\rm max}$ increases; see Sec.~\ref{section:Regularity_Of_Seff} for details.}
\label{fig:Seff_nmax_m0_r010}
\end{figure*}

In Fig.~\ref{fig:Seff_nmax_m0_r010}, we plot the effective source $S^{\rm eff}_{m=0}$ for different puncture orders $\cn_\text{max}$ with $r_p/M=10.0$.
For this plot, we applied an analytical mesh refinement, in the style of \cite{Macedo22}, around the lines $r=r_p$ and $\theta=\pi/2$ in order to show more clearly the behavior of the effective source there. We can confirm by eye that the effective source is $C^0$ at the particle for $\cn_\text{max}=1$, as predicted, and that it becomes increasingly regular there as the order of the puncture is increased. 

We remark here that the overall shape of the effective source changes rather drastically from order to order. In particular, the magnitude of the contributions for increasing $\cn$ does not significantly decrease as one might naively expect, even in a neighbourhood of $(r=r_p,\theta=\pi/2)$. The reason is that the puncture is obtained as a local expansion near the particle, while the individual $m$-modes are sensitive to the puncture's behavior far from the particle: when integrating over $S^1$ to obtain the $m$-modes, one integrates over a distant region on the opposite side of the black hole, and the puncture's behavior there contributes to each $m$-mode even at $(r=r_p,\theta=\pi/2)$. In principle, one can freely choose how to smoothly extend the puncture outside the local neighbourhood, altering the individual modes. In the present case, we simply chose the trivial analytic extension outside the normal neighborhood, whereby we use the same expansion in powers of the comoving coordinates $X^i$ for the full range $\varphi-\Omega_p t \in [-\pi, \pi]$. As emphasised here, this choice of extension, combined with our particular choice of  $X^i$, has the major benefit of making the $m$-mode integrals analytically tractable.

\section{Conclusion}
\label{sec:conclusion}

In this paper, we have computed the singular field for a scalar charge to $14$th order in distance to the particle, $10$ orders higher than the previous state of the art. There is no obstacle in going to higher orders, but as the companion paper shows~\cite{paperII}, doing so would have diminishing returns.

Unlike traditional methods, we use a very simple conceptual framework: adopting a local coordinate frame around the particle, writing the field equations as a Poisson equation plus small corrections, and solving it by decomposing the equations of motion in spherical harmonics centered around the particle. Besides its practical utility, our approach avoids the need for technical tools that might be unfamiliar outside the self-force community. Our analysis in Sec.~\ref{local expansion of PhiS} also highlights that the Detweiler-Whiting regular field is actually identical (in a local neighborhood of the particle) to a Hadamard \emph{partie finie} regularization of the physical, retarded field, which might again make it more accessible to the intuition of post-Newtonian experts, for example.

However, we do view our approach primarily as having practical, rather than conceptual advantages. If the local coordinates are well chosen (to have a simple relationship with the global, inertial coordinates used to solve the field equations), our procedure yields the puncture directly in a user-ready form adapted to one's problem. Furthermore, we have shown that one can analytically compute the $m$-modes of the puncture around the central black hole, at all orders in distance.  This ability to construct exact analytical puncture modes (also previously shown in Ref.~\cite{Thornburg:2016msc}) represents a potentially large advantage for $m$-mode puncture schemes over $\ell m$-mode ones. As discussed in Refs.~\cite{Miller16,Toomani2022,Bourg:2024vre}, traditional methods of obtaining $\ell m$ modes of a puncture involve applying additional approximations that (i) lead to large-$\ell$ divergences at points away from the particle ($\Delta r\neq0$), and (ii) prevent one from applying Ref.~\cite{Miller16}'s strategy for constructing the quadratic source term in the second-order Einstein equations. Current second-order calculations~\cite{Pound:2019lzj,Warburton:2021kwk,Wardell:2021fyy} and the puncture scheme in Ref.~\cite{Bourg:2024vre} overcome this by numerically evaluating the integral over the sphere to obtain the puncture's $\ell m$ modes. More concretely, they analytically evaluate the integral over one of the angles and numerically evaluate the integral over the second angle. As reviewed in Sec.~\ref{sec:advantages}, generating current second-order results in this way involved the evaluation of hundreds of millions of numerical integrals, which an $m$-mode scheme would bypass. Of course, this benefit comes at the cost of introducing more burdensome numerics to solve PDEs in $(r,\theta)$.

Our particular form of the puncture might also make the $\ell$-mode integrals more analytically tractable, but we leave exploration of that possibility to future work.


Interestingly, the overall size of the expressions for the modes of the singular field, when centered around the particle, only increases \textit{linearly} with the order of the puncture. In particular, evaluating the four-dimensional puncture, even for a $14$th-order puncture is very fast. Unfortunately, when decomposing into modes on spheres or rings centered around the central black hole, the overall size of the expressions increases dramatically with the order of the puncture. In particular, the results of this paper highlight an interesting tension: On the one hand, it is far more natural to decompose the solution into modes on the 2-sphere centered around the particle instead of the central black hole. There are two ways to intuitively understand this. First, these 2-spheres, in some intuitive sense, only capture the local behaviour of the singular field. When decomposing around the central black hole for example, one needs to choose an extension of the puncture beyond the particle's neighbourhood so that it is defined on the entire 2-sphere around the central black hole. Second, the 2-sphere around the particle never intersects the particle itself, meaning that the modes always describe a smooth function.
On the other hand, the field equations are more naturally written in coordinates centered around the central black hole, in order to take advantage of the symmetries of the background metric. Much of the current self-force program has favoured the latter aspect. In hindsight, this has come at a hefty price for anything related to the puncture: analytical expressions are difficult to obtain beyond a few orders and one is plagued with poor $\ell$-sum convergence even far from the particle. This paper highlights that it may be worthwhile to instead consider a coordinate system and spherical decomposition that caters to the puncture's needs, instead of the symmetries of the background metric, at least when it comes to solving the field equations in the vicinity of the particle.

We foresee a variety of applications of the high-order punctures that can be obtained with our method. 
In a puncture scheme, the residual field is the numerical variable, and its degree of smoothness is directly linked with the order of the puncture: the higher the order, the smoother the residual field. As demonstrated in our analysis of $m$-mode convergence, increased smoothness translates into tangibly faster convergence properties of numerical algorithms. Some possible concrete applications of this are the following: 
\begin{enumerate}
    \item If using a spectral method in an $m$-mode puncture scheme, finite differentiability of the effective source will lead to slow convergence of the spectral solver. A higher-order puncture, and therefore a smoother effective source, ameliorates this problem. The same considerations apply to a finite-difference $m$-mode scheme, which must use a finer grid around the particle's position to resolve the features of the effective source there.
    \item In second-order self-force calculations, a higher-order puncture would lead to more rapid convergence of the quadratic source term, when employing the strategy in Ref.~\cite{Miller16}. It could also lead to a reduction in the number of radial grid points needed to resolve the source around the particle.
    \item In frequency-domain methods for eccentric orbits, the particle singularity leads to  Gibbs phenomena~\cite{Barack:2008ms,Hopper:2012ty}. The method of extended homogeneous solutions, designed to overcome this problem~\cite{Barack:2008ms,Hopper:2012ty}, can require extremely high precision to cancel large numbers~\cite{vandeMeent:2017bcc}. A sufficiently smooth effective source could make the Gibbs phenomena negligible, avoiding the need for extended homogeneous solutions. This would provide an alternative to the approach in Ref.~\cite{Leather:2023dzj}.
    \item The Teukolsky puncture scheme in Ref.~\cite{Bourg:2024vre} encounters complicated $\ell$-mode behavior before reaching the expected exponential convergence away from the particle. Working with a smoother effective source could bypass this behavior.
    \item The recently developed Lorenz-gauge metric reconstruction method~\cite{Dolan:2023enf} can involve large cancellations between different contributions to the solution. A high-order puncture scheme should avoid this issue by making the source term for each contribution highly differentiable.
\end{enumerate}

Realising these potential benefits will require applying our method to the gravitational case and to more generic scenarios: a Kerr background and eccentric or inclined orbits. As a first step, work is underway to extend our results to the case of gravity for circular orbits on a Kerr background. Since the the general method we adopt (from Ref.~\cite{Pound:2012nt}) is valid for gravity in a generic spacetime, the only technical requirement lies in constructing a suitable local coordinate system. For circular orbits in Kerr, one can easily find a natural analogue of the coordinate system  we used in this paper. The extension to eccentric or inclined orbits will be more challenging.


Of course, from a practical point of view, a certain balance that must be struck: a high-order puncture results in faster convergence properties, but computing the puncture and effective source  takes longer. This is explored in the companion paper~\cite{paperII}. We also note that the potential advantages we have listed are features of puncture schemes with sufficiently high-order punctures, as opposed to being advantages of our particular method of calculating those punctures. At first order in self-force theory, traditional methods are almost certainly able to calculate punctures to sufficiently high order in distance to realise the benefits we have suggested; the calculations in Ref.~\cite{Heffernan12}, for example, did not face fundamental obstacles in proceeding to higher orders (though we have found in practice that our new method is more computationally efficient). We also note that for most first-order self-force calculations, mode-sum regularization~\cite{Barack:2009ux} has been favored over puncture schemes because it is generally more efficient and often simpler to implement. Our enumerated list above proposes some cases in which a puncture scheme, with a sufficiently high order puncture, should be advantageous, but mode-sum regularization will certainly remain the preferred option in many (or even most) first-order self-force computations.



\begin{acknowledgments}

We thank Barry Wardell and Leor Barack for helpful discussions. 
PB acknowledges the support of an EPSRC Fellowship in Mathematical Sciences and of the Dutch Research Council (NWO) (project name: Resonating with the new gravitational-wave era, project number: OCENW.M.21.119). 
AP acknowledges the support of a Royal Society University Research Fellowship, and AP and SDU acknowledge support from a Royal Society Research Grant for Research Fellows and from the ERC Consolidator/UKRI Frontier Research Grant GWModels (selected by the ERC and funded by UKRI grant number EP/Y008251/1). 
AP and PB additionally acknowledge the support of a Royal Society University Research Fellowship Enhancement Award. 
SDU additionally acknowledges the support of the fellowship Lumina Quaeruntur No.~LQ100032102 of the Czech Academy of Sciences. 
RPM acknowledges support from the Villum Investigator program supported by the VILLUM Foundation (grant no. VIL37766) and the DNRF Chair program (grant no. DNRF162) by the Danish National Research Foundation and the European Union's Horizon 2020 research and innovation programme under the Marie Sklodowska-Curie grant agreement No 101131233.  RPM is a long-term research visitor at the STAG Research Centre, University of Southampton and this project has also received funding from the STFC Grant No.~ST/V000551/1.
\end{acknowledgments}

\appendix

\section{Computing \texorpdfstring{$\cC_{\cl \cm \cn}$}{C̅{\_}l̅ m̅ n}} \label{sec:clmn}

In this appendix we illustrate how to calculate the mode coefficients in Eq.~\eqref{CTerm}, which appear in the puncture modes through Eq.~\eqref{PhiPlmnClmn}.

\subsection{Reduction to a general integral}

Note that the metric components $g_{\alpha\beta}$ are smooth functions of the comoving coordinates $X^i=(X,Y,Z)$. As a result, they can be immediately expanded in a Taylor series in terms of the unit vector $N^i \coloneqq \bracket{\frac{X}{R},\frac{Y}{R},\frac{Z}{R}}$,
\begin{equation}
	g_{\alpha\beta} = \sum_{\cn\geq0} g_{\alpha\beta \,i_1\cdots i_\cn} R^n N^{i_1}\cdots N^{i_\cn},
\end{equation}
in analogy with Eq.~\eqref{PhiR inertial coordinate expansion}. Note that this is \textit{not} an STF decomposition.

On the other hand, the scalar field is expanded as in Eq.~\eqref{phi_expansion}. Since $\PhiP$ is real, we can re-write the expansion so that it appears manifestly real:
\begin{align}
	\PhiP &= \sumnFull \suml \sum_{\cm=0}^{\cl} \frac{\PhiP_{\cl \cm \cn}}{1+\delta_0^\cm} \Ymod(\ctheta,\cphi), \\
	\Ymod(\ctheta,\cphi) &\coloneqq 2 \Ylm(\ctheta,0) \cos (\cm \cphi).
\end{align}
Note that the summation is over non-negative $m$-modes. The negative $m$-modes of $\PhiP $ are obtained via the reality condition~\eqref{PhiP -m real}.

The expression for $\hat{C}(\Phi)$ will therefore be a sum over product of spherical harmonics and $N^i$'s. In order to cast it into the form \eqref{CTerm}, we need to re-express it as a sum over spherical harmonics. This implies the computation of the following general integral:
\begin{align}
	\MoveEqLeft[4] \mathcal{I}_{\ell_1 m_1}^{\ell_2 m_2}(t_1,t_2,t_3,t_4;a,b) \\
	\coloneqq&{} \int_{S^2} \cos^{t_1} \varphi \sin^{t_2} \varphi \cos^{t_3} \theta \sin^{t_4} \theta \nonumber \\
	   & \times \bracket{\partial_\theta^a \partial_\varphi^b \tilde{Y}_{\ell_1 m_1}(\theta, \varphi)} \bracket{Y_{\ell_2 m_2}}^\star(\theta, \varphi) d\Omega,
\end{align}
where $a,b = 0,1,2$, $d\Omega = \sin \theta \, d \theta d \varphi$, $t_i \geq 0$, except for $t_4$ which may be negative. In the above, we dropped the bars for convenience. 

We show next that $\mathcal{I}$ can be re-written explicitly as sums over quantities depending on the $t_i$, $a$, $b$, $\ell_j$, and $m_k$.
To facilitate this, we split $\mathcal{I}$ into its $\theta$ and $\varphi$ integral,
\begin{multline}
    \mathcal{I}_{\ell_1 m_1}^{\ell_2 m_2}(t_1,t_2,t_3,t_4;a,b) \\
        = c_{\ell_1 m_1} c_{\ell_2 m_2} \Theta_{\ell_1 m_1}^{\ell_2 m_2}(t_3,t_4;a) \Psi_{m_1}^{m_2}(t_1,t_2;b), \label{eqn:IThetaPsi}
\end{multline}
where
\begin{align}
    \Theta&_{\ell_1 m_1}^{\ell_2 m_2}(t_3,t_4;a) \coloneqq \nonumber \\
        & \int_0^\pi \cos^{t_3} \theta \sin^{t_4+1} \theta \,\partial_\theta^a \bracket{P_{\ell_1}^{m_1}(\cos \theta)} P_{\ell_2}^{m_2}(\cos \theta) d\theta,\\
    \Psi&_{m_1}^{m_2}(t_1,t_2;b) \coloneqq \nonumber \\
        & 2 \int_0^{2 \pi} \cos^{t_1} \varphi \sin^{t_2} \varphi \, \partial_\varphi^b \bracket{\cos (m_1 \varphi)} e^{-i m_2 \varphi} d \varphi, \\
    c_{\ell m} &\coloneqq \sqrt{\frac{2\ell+1}{4 \pi} \frac{(\ell-m)!}{(\ell+m)!}}.\label{clm}
\end{align}
The next two subsections deal with each of the two separated integrals in turn.

\subsection{Evaluation of the \texorpdfstring{$\varphi$}{φ} integral}

We first work out an expression for $\Psi$.
Expanding $\cos m_1 \varphi$ in terms of exponentials, we put $\Psi$ in the form
\begin{align}
\Psi_{m_1}^{m_2}(t_1,t_2;b) ={}& (i m_1)^b \left[\Psibar(t_1,t_2,m_1-m_2)\right. \nonumber \\
    & \left. + (-1)^b \Psibar(t_1,t_2,-m_1-m_2) \right], \label{eqn:Psi_Psibar}\\
\Psibar(t_1,t_2,m) \coloneqq{}& \int_0^{2 \pi} \cos^{t_1} \varphi \sin^{t_2} \varphi e^{i m \varphi} d \varphi.
\end{align}
$\Psibar$ is computed by writing the remaining trigonometric functions in terms of complex exponentials and using the binomial theorem, as well as the identity
\begin{equation}
\int_0^{2 \pi} e^{im\varphi} d\varphi = 2 \pi \delta_0^m.
\end{equation}
We find
\begin{multline}
\Psibar(t_1,t_2,m) \\
    = \frac{2\pi i^{t_2}}{2^{t_1+t_2}} \sum_{k_1=0}^{t_1} \sum_{k_2=0}^{t_2} (-1)^{k_2} \binom{t_1}{k_1}\binom{t_2}{k_2} \delta_{t_1+t_2-m}^{2 (k_1+k_2)}. \label{eqn:Psibar}
\end{multline}
This formula gives rise to the following selection rules:
\begin{itemize}
\item $\Psibar=0$ if any of the following conditions hold:
\begin{enumerate}
	\item $t_1+t_2-m$ is odd,
	\item $t_1+t_2<|m|$.
\end{enumerate}
\end{itemize}

\subsection{Evaluation of the \texorpdfstring{$\theta$}{θ} integral}

Turning our attention now to $\Theta$, let us first consider the case where $a=0$.
Making the change of variable $x = \cos \theta$, we write
\begin{multline}
    \Theta_{\ell_1 m_1}^{\ell_2 m_2}(t_3,t_4;0) \\
        = \int_{-1}^1 x^{t_3} \bracket{\sqrt{1-x^2}}^{t_4} P_{\ell_1}^{m_1}(x) P_{\ell_2}^{m_2}(x) dx. \label{eqn:Theta_a0_formula}
\end{multline}

In order to make progress, we reduce $\Theta$ to the case where $t_3=t_4=0$.

We can first reduce $\Theta$ to the case $t_3=0$, by repeatedly making use of the following relation:
\begin{multline}
    (2\ell+1) x \Plm(x) = (\ell+m) P_{\ell-1}^m(x) \\
        + (\ell-m+1) P_{\ell+1}^m(x). \label{eqn:Legendre_property1}
\end{multline}
The case for $t_4$ is more subtle, since, due to the use of spherical coordinates, some integrals have $t_4 < 0$. Depending on the sign of $t_4$, we use the two identities
\begin{align}
(2\ell+1) \sqrt{1-x^2} \Plm(x) &= P_{\ell-1}^{m+1}(x) - P_{\ell+1}^{m+1}(x), 
\label{eqn:Legendre_property2} \\
\frac{-2m}{\sqrt{1-x^2}} \Plm(x) &= P_{\ell-1}^{m+1}(x) +\nonumber \\
&(\ell+m-1)(\ell+m) P_{\ell-1}^{m-1}(x). \label{eqn:Legendre_property3}
\end{align}
Note that the last identity assumes that $m \neq 0$. In the computation of $\Theta$, we then make use of both $P_{\ell_1}^{m_1}$ and $P_{\ell_2}^{m_2}$ in order to decrease (or increase) $t_4$ to zero.
The only case where this recursive procedure does not work is when $t_4 < 0$ and both $m_1=m_2=0$, as one cannot use \eqref{eqn:Legendre_property3} on either of the Legendre polynomials. However, we can calculate this specific case using the series representations of the Legendre polynomials:
\begin{align}
\MoveEqLeft[2] \Theta_{\ell_1 0}^{\ell_2 0}(0,t_4;0) = \int_{-1}^1 \bracket{\sqrt{1-x^2}}^{t_4} P_{\ell_1} P_{\ell_2} dx \nonumber \\
    ={}& 2^{\ell_1+\ell_2}\int_{-1}^{1}\biggl(\sum_{j=0}^{\ell_1}a_{\ell_{1}j}x^j\biggr)\biggl(\sum_{k=0}^{\ell_2}a_{\ell_{2}k}x^k\biggr) (1-x^2)^{\frac{t_{4}}{2}} dx\nonumber \\
    ={}& 2^{\ell_1+\ell_2}\sum_{k=0}^{\ell_1+\ell_2}b_{\ell_{1}\ell_{2}k}\int_{-1}^{1} (1-x^2)^{\frac{t_{4}}{2}}x^k dx \nonumber \\
    ={}& 2^{\ell_1+\ell_2-1}\sum_{k=0}^{\ell_1+\ell_2}b_{\ell_{1}\ell_{2}k}\left(1+(-1)^k\right)B\left(\frac{1+k}{2},1+\frac{t_{4}}{2}\right), \label{eqn:Theta_specialCase}
\end{align}
where $B(x,y)$ is the standard beta function, \(a_{nk}=\binom{n}{k}\binom{\frac{k+n-1}{2}}{n}\), and
\begin{align}
    b_{\ell_1\ell_2 k} ={}& \sum^k_{j=0}a_{\ell_{1}j}a_{\ell_{2},k-j} \nonumber \\
        ={}& \sum^k_{j=0}\binom{\ell_1}{j}\binom{\frac{\ell_1+j-1}{2}}{\ell_1}\binom{\ell_2}{k-j}\binom{\frac{\ell_{2}+k-j-1}{2}}{\ell_{2}}.
\end{align}
In the second line, we have used the series representation of the Legendre polynomials and to go to the third line, we take the Cauchy product of the two series. The integral can then directly be evaluated using Eq.~(3.251.1) of Ref.~\cite{Grad14}.
Note that the result is finite if and only if $t_4 \geq -1$. With the exception of this particular case, we can also reduce to the case $t_4=0$. We are then left with the case $t_3=0=t_4$, which corresponds to the integral over a product of two Legendre polynomials. 


An explicit formula exists for this case~\cite{Caola78,Wong98},
\begin{align}
&\Theta_{\ell_1 m_1}^{\ell_2 m_2}(0,0;0) = \frac{1 + (-1)^{\ell_1+\ell_2-m_1-m_2}}{2} \times \nonumber \\
&\sum_{k_1=0}^{\floor{\frac{\ell_1-m_1}{2}}} \sum_{k_2=0}^{\floor{\frac{\ell_2-m_2}{2}}} c_{\ell_1 m_1 k_1} c_{\ell_2 m_2 k_2} \times \nonumber \\
&B\left(\frac{\ell_1+\ell_2-m_1-m_2-2k_1-2k_2+1}{2},  \right. \nonumber \\
&\left. \frac{m_1+m_2+2k_1+2k_2+2}{2} \right),
\label{eqn:Theta_baseCase}
\end{align}
where $\floor{x}$ denotes the integer part of $x$ and
\begin{equation}
c_{\ell mp} \coloneqq (-1)^{m+p} \frac{(\ell+m)!}{2^{m+2p} (m+p)! p! (\ell-m-2p)!}.
\end{equation}
Note that in the above, our definition of $c_{\ell mp}$ differs from the one in Ref.~\cite{Caola78} by a factor $(-1)^m$ to reflect a difference in notation in the Legendre polynomial. This expression is only valid for $\ell$, $m \geq 0$.
However, as a result of the recursive algorithm, some of the integrals will feature Legendre polynomials where $\ell<0$ and/or $m<0$. For those integrals, before using the formula \eqref{eqn:Theta_baseCase}, one first uses the following properties of the Legendre polynomials:
\begin{align}
P_\ell^{-m} &= (-1)^m \frac{(\ell-m)!}{(\ell+m)!} \Plm, \\
P_{(-\ell)}^m &\coloneqq P_{(\ell-1)}^m.
\end{align}
Note that we have the following  selection rule: $\Theta=0$ if $\ell_1+\ell_2-m_1-m_2$ is odd.

Finally, we can comment on the case for $a=1$ or $2$.
$\partial_\theta \Plm$ can be expressed as a sum of two Legendre polynomials,
\begin{multline}
    \partial_\theta \Plm(\cos \theta) = m \cot \theta \Plm(\cos \theta) \\
        +  \frac{\sin \theta}{|\sin \theta|} P_\ell^{m+1}(\cos \theta).
\end{multline}
We can then relate the case the integral for $a=1$ in terms of (linear combinations of) integrals with $a=0$.
Since the range of integration for $\theta$ is $\theta \in [0,\pi]$, we have
\begin{multline}
    \Theta_{\ell_1 m_1}^{\ell_2 m_2}(t_3,t_4;1) = m \Theta_{\ell_1 m_1}^{\ell_2 m_2}(t_3+1,t_4-1;0) \\
        + \Theta_{\ell_1 (m_1+1)}^{\ell_2 m_2}(t_3,t_4;0). \label{eqn:Theta_a1_formula}
\end{multline}

In a similar fashion, $\partial_\theta^2 \Plm$ is written as the sum of three Legendre polynomials, which implies
\begin{align}
    \MoveEqLeft[4] \Theta_{\ell_1 m_1}^{\ell_2 m_2}(t_3,t_4;2) \nonumber \\
        ={}& m_1(m_1-1) \Theta_{\ell_1 m_1}^{\ell_2 m_2}(t_3+2,t_4-2;0) \nonumber \\
        & + (2m_1+1) \Theta_{\ell_1 (m_1+1)}^{\ell_2 m_2}(t_3+1,t_4-1;0) \nonumber \\
        & - m_1 \Theta_{\ell_1 m_1}^{\ell_2 m_2}(t_3,t_4;0) + \Theta_{\ell_1 (m_1+2)}^{\ell_2 m_2}(t_3,t_4;0). \label{eqn:Theta_a2_formula}
\end{align}

In summary, the general integral $\mathcal{I}_{\ell_1 m_1}^{\ell_2 m_2}(t_1,t_2,t_3,t_4;a,b)$ is decomposed into a $\varphi$ and $\theta$-integral, Eq.~\eqref{eqn:IThetaPsi}. The $\varphi$-integral is given via Eq.~\eqref{eqn:Psi_Psibar} and \eqref{eqn:Psibar}.
The $\theta$ integral, $\Theta_{\ell_1 m_1}^{\ell_2 m_2}(t_3,t_4;a)$, is computed through the following steps:
\begin{enumerate}
    \item If $a=1$ or $2$, we can use Eq.~\eqref{eqn:Theta_a1_formula} and Eq.~\eqref{eqn:Theta_a2_formula} respectively to reduce it to the case $a=0$.
    \item The integral for $a=0$ is given in Eq.~\eqref{eqn:Theta_a0_formula}. This integral is solved by noticing that the case $t_3=t_4=0$ has a known analytical solution, given in Eq.~\eqref{eqn:Theta_baseCase}.
    \begin{enumerate}
        \item First, apply Eq.~\eqref{eqn:Legendre_property1} repeatedly to reduce $t_3$ to $0$.
        \item Then, use the identities given in Eq.~\eqref{eqn:Legendre_property2} and \eqref{eqn:Legendre_property3} repeatedly to reduce $t_4$ to $0$ as well. These identities only work for $m \neq 0$. As a result, in the event where both $m_1 = m_2 = 0$, but $t_4 \neq 0$, these relations cannot be used to reduce $t_4$ to $0$. For that special case, however, the integral can be evaluated explicitly; see Eq.~\eqref{eqn:Theta_specialCase}.
    \end{enumerate}
\end{enumerate}


\section{The effective source \texorpdfstring{$S^\text{eff}_m$}{S{\textasciicircum}eff{\_}m}}\label{app:Seffm}

Given an $m$-mode of the puncture field, we can immediately compute the corresponding $m$-mode of the effective source~\eqref{ResFieldEqnTMP} as
\begin{align}
	S^\text{eff}_m &= -\square_m \PhiP_m \nonumber \\
	&=  -\sum_{\cn=-1}^{\cn_\text{max}} \sum_{\cl=0}^{3\cn+3} \sum_{\cm=0}^\cl \frac{\Phi_{\cl\cm\cn} c_{\cl\cm}}{\pi (1+\delta_0^m)} \square_m \bigl[\cos(\cm\cphi) \Ilmnm(\varrho)\bigr],
\end{align}
where $\square_m$ is given in \eqref{eqn:square0}--\eqref{eqn:square2}.
Because we use recursion relations to obtain the analytical expressions for $\PhiP_m$ and $\Ilmnm(\varrho)$, it is more efficient to derive how the operator $\square$ modifies the recursion relation, rather than applying $\square$ to explicit functions of $(r,y)$.
In the following, we will suppress the indices in the expression $\Ilmnm$, simply writing $\mathfrak{I}$ instead.

\subsection{Contribution of \texorpdfstring{${}^0\square_m$}{{\textasciicircum}0 □{\_}m}}

${}^0\square_m$ simply contributes to a factor of the singular field,
\begin{equation}
	{}^0\square_m \PhiP_m = m^2 \bracket{\frac{\Omega_p^2}{f} - \frac{1}{r^2 (1-y)}} \PhiP_m. \label{Square0CosContribution}
\end{equation}

\subsection{Contribution of \texorpdfstring{${}^1\square$}{{\textasciicircum}1 □}}

Since ${}^1\square$ is a first-order linear operator, we can apply the chain rule:
\begin{equation}
	{}^1\square\left[\cos(\cm \cphi) \mathfrak{I}\right] = \left[{}^1\square\cos(\cm \cphi)\right] \mathfrak{I}+ \cos(\cm \cphi)\; {}^1\square \mathfrak{I},
\end{equation}
where
\begingroup
\allowdisplaybreaks
\begin{align}
	\MoveEqLeft[3] {}^1\square \cos(\cm \cphi) \nonumber\\*
        ={}& \frac{1}{r^2}\left[\bracket{r_p+f_p r_p +5 \Delta r}-\frac{\Delta r}{y}\right] \frac{\partial}{\partial r} \cos(\cm \cphi), \label{Square1CosContribution} \\
	\MoveEqLeft[3] \frac{\partial}{\partial r} \cos(\cm \cphi) \nonumber\\*
        ={}& -\frac{\cm r_p}{\varrho^2} \sqrt{\frac{y}{f_p}} \sin \left[\cm \arctan \bracket{\Delta r, -r_p \sqrt{f_p y}}\right], \\
	{}^1\square \mathfrak{I} ={}& \left[\frac{r_p + f_p r_p + 2\Delta r}{f_p} \Delta r + (1-3y) r_p^2\right] \frac{\mathfrak{I}'}{\varrho r^2},\label{Square1IContribution}
\end{align}%
\endgroup
where we defined the quantity $\mathfrak{I}' \coloneqq \partial_\varrho \mathfrak{I}$. We return to how this quantity is computed at the end of this Appendix. In the above, the arctan function with two arguments is defined in the interval $-\pi < \arctan(x,y) \leq \pi$ and returns the principal value of the argument of $x+iy$.

We remark that the right-hand side of \eqref{Square1CosContribution} is well defined in the limit $y \to 0$.
Specifically,
\begin{equation}
	\lim\limits_{y \to 0^+} \frac{1}{y} \frac{\partial}{\partial r} \cos(\cm \cphi) = \frac{f_p \cm^2 r_p^2}{\Delta r^3}
	\begin{cases} 
		1 & \Delta r > 0, \\
		\cos (\cm \pi) & \Delta r < 0.
	\end{cases}
\end{equation}
The factor of $\cos (\cm \pi)$ for $\Delta r < 0$ arises from the fact that the argument of $\arctan(\cdot)$ is defined on the interval $[-\pi/2,\pi,2]$ and a correction of $+\pi$ must be added to lie in the correct quadrant.

\subsection{Contribution of \texorpdfstring{${}^2\square$}{{\textasciicircum}2 □}}

Since ${}^2\square$ only contains second-order differential operators, the chain rule implies
\begin{multline}
	{}^2\square[\cos(\cm \cphi) \mathfrak{I}] = \left[{}^2\square\cos(\cm \cphi)\right] \mathfrak{I} \\+ \cos(\cm \cphi)\; {}^2\square\mathfrak{I} + CT,
\end{multline}
where
\begin{align}
	{}^2\square&\cos(\cm \cphi) = -\frac{\cm r_p}{r^2  \varrho^4 \sqrt{f_p^3 y}} \times \nonumber \\
	&\Bigg[ \cm r_p \sqrt{f_p y} \Big\{ \vphantom{\sqrt{x}} r_p f_p r y + \Delta r \bracket{r_p y+\Delta r} \Big\}  \cos(\cm \cphi)  \nonumber \\
	& \quad +\Delta r \Big\{ f_p r_p y (r_p-2 \Delta r-3 r_p y)   \nonumber \\
	& +\Delta r \bracket{\Delta r-2r_p y-3 y \Delta r} \Big\}  \sin(\cm \cphi) \Bigg], 
	\label{Square2CosContribution} \\
	{}^2\square&\mathfrak{I} =\frac{f_p^2 r_p^4 y (1-y)+r \Delta r^2 (f_p r_p + \Delta r)}{f_p^2 r^2 \varrho^2} \mathfrak{I}'' \nonumber \\
	& +\frac{-f_p^2 r_p^4 y (1-y) + r (f_p r_p+\Delta r) (f_p \varrho^2-\Delta r^2)}{f_p^2 r^2 \varrho^3} \mathfrak{I}', \label{Square2IContribution} \\
	CT &= \biggl[2 f \frac{\partial}{\partial \Delta r} \cos(\cm \cphi) \frac{\partial}{\partial \Delta r} \mathfrak{I} \nonumber\\
 &\qquad + \frac{8y(1-y)}{r^2} \frac{\partial}{\partial y} \cos(\cm \cphi) \frac{\partial}{\partial y} \mathfrak{I}\biggr]. \label{Square2Expr}
\end{align}
In the above, we defined $\mathfrak{I}'' := \partial_{\varrho}^2 \mathfrak{I}$.
The cross term $CT$ in \eqref{Square2Expr} can be re-written as
\begin{equation}
	CT \coloneqq 2 \left[\frac{f}{f_p}-\frac{r_p^2}{r}(1-y)\right] \frac{\Delta r}{\varrho} \mathfrak{I}' \frac{\partial}{\partial r} \cos(\cm \cphi). \label{CrossTermContribution}
\end{equation}

Putting it all together, we can write the effective source as
\begin{equation}
	S^\text{eff}_m = -\sum_{\cn=-1}^{\cn_\text{max}} \sum_{\cl=0}^{3\cn+3} \sum_{\cm=0}^\cl \frac{\Phi_{\cl\cm\cn} c_{\cl\cm}}{\pi (1+\delta_0^m)} \bracket{\mathcal{A} \mathfrak{I}+\mathcal{B} \mathfrak{I}'+\mathcal{C} \mathfrak{I}''},
\end{equation}
where
\begin{align}
	\mathcal{A} &\coloneqq \square_m \cos(\cm \cphi) \nonumber\\
 &\hphantom{:}= \underbrace{{}^0\square_m \cos(\cm \cphi)}_\text{\eqref{Square0CosContribution}}+\underbrace{{}^1\square \cos(\cm \cphi)}_\text{\eqref{Square1CosContribution}}+\underbrace{{}^2\square \cos(\cm \cphi)}_\text{\eqref{Square2CosContribution}}, \\
	\mathcal{B} &\coloneqq \cos (\cm \cphi) \underbrace{\frac{{}^1\square\mathfrak{I}}{\mathfrak{I}'}}_\text{\eqref{Square1IContribution}} + \underbrace{\frac{CT}{\mathfrak{I}'}}_\text{\eqref{CrossTermContribution}} + \underbrace{\#_1}_\text{\eqref{Square2IContribution}} \cos (\cm \cphi),\\
	\mathcal{C} &\coloneqq \underbrace{\#_2}_\text{\eqref{Square2IContribution}} \cos (\cm \cphi).
\end{align}
In the above, $\#_1$ and $\#_2$ stand respectively for the coefficients in front of $\mathfrak{I}'$ and $\mathfrak{I}''$ in \eqref{Square2IContribution}.

In order to complete the calculation, we discuss how 
$\mathfrak{I}'$ and $\mathfrak{I}''$ are computed.
First, note that $\mathfrak{I}'$ and $\mathfrak{I}''$ obey the same recursion relation as $\mathfrak{I}$; see  Eq.~\eqref{recursion_relation}. The only modifications in the algorithm are: (i) the formula for the ``top case'' Eq.~\eqref{top case 1}, (ii) the ``base case'', Eq.~\eqref{I00nm}. In principle, it is therefore sufficient to differentiate both equations with respect to $\varrho$ and use the resulting expressions for $\partial_{\varrho} \mathfrak{I}^m_{\cl \cl \cn}$ and $\partial_{\varrho}^2 \mathfrak{I}^m_{\cl \cl \cn}$ to close the recursive algorithm. However, in the interest of speeding up the calculation, it turns out that the latter two expressions can be related to $\mathfrak{I}^m_{\cl \cm \cn}$. Specifically, from the integral representation of $\mathfrak{I}^m_{00 \cn}$, Eq.~\eqref{I00nmIntegral}, we can deduce the relation
\begin{equation}
    \partial_{\varrho} \mathfrak{I}^m_{00 \cn} = n \varrho \mathfrak{I}^m_{0,0,\cn-2}.
    \label{eqn:dI00nm}
\end{equation}
As a result, differentiating the top-case \eqref{top case 1}, we find that
\begin{equation}
    \varrho \partial_{\varrho} \mathfrak{I}^m_{\cl \cl \cn} = \cl \, \mathfrak{I}^m_{\cl \cl \cn} + \frac{\cn-\cl}{(2\cl+1) (2\cl+3)} \mathfrak{I}^m_{\cl+2, \cl+2, \cn},
\end{equation}
from which Eq.~\eqref{eqn:dI00nm} can be treated as a special case. Differentiating this formula, we obtain
\begin{align}
    \varrho^2 \partial_{\varrho}^2 \mathfrak{I}^m_{\cl \cl \cn} &= \cl (\cl-1) \, \mathfrak{I}^m_{\cl \cl \cn} + \frac{\cn-\cl}{2\cl+3} \mathfrak{I}^m_{\cl+2, \cl+2, \cn} \nonumber \\
    &+ \frac{(\cn-\cl) (\cn-\cl-2)}{(2\cl+1) (2\cl+3) (2\cl+5) (2\cl+7)} \mathfrak{I}^m_{\cl+4, \cl+4, \cn}.
\end{align}

\section{Mode coefficients \texorpdfstring{$\PhiP_{\cl, \cm, \cn}$}{Φ{\textasciicircum}P{\_}l̅,m̅,n}}
\label{sec:PhilmnExpr}

We give below all the non-zero mode coefficients $\PhiP_{\cl, \cm, \cn}$ for $\cn=-1$, $0$, $1$, and $2$. By virtue of Eq.~\eqref{PhiP -m real}, we here only give the modes for $\cm \geq 0$.

\begin{subequations}
\begin{align}
     \PhiP_{0, 0, -1} &= 2 \sqrt{\pi} q.
\end{align}
\end{subequations}

\begin{subequations}
\begin{align}
     \PhiP_{1, 1, 0} &= \frac{\sqrt{\frac{\pi}{6}} (3 f_p-1) q}{5 \sqrt{f_p} r_p}, \\
     \PhiP_{3, 1, 0} &= -\frac{\sqrt{\frac{\pi}{21}} (f_p-7) q}{20 \sqrt{f_p} r_p}, \\
     \PhiP_{3, 3, 0} &= -\frac{\sqrt{\frac{\pi}{35}} (f_p+1) q}{4 \sqrt{f_p} r_p}.
\end{align}
\end{subequations}

\begingroup
\allowdisplaybreaks
\begin{subequations}
\begin{align}
     \PhiP_{0, 0, 1} &= \frac{\sqrt{\pi } \left(120 f_p^2-69 f_p-17\right) q}{840 f_p r_p^2}, \\
     \PhiP_{2, 0, 1} &= \frac{\sqrt{\frac{\pi }{5}} \left(3 f_p^2-3 f_p+4\right) q}{42 f_p r_p^2}, \\
     \PhiP_{2, 2, 1} &= \frac{\sqrt{\frac{\pi }{30}} \left(2 f_p^2-f_p+1\right) q}{7 f_p (3 f_p-1) r_p^2}, \\
     \PhiP_{4, 0, 1} &= -\frac{\sqrt{\pi } \left(10 f_p^2+39 f_p-3\right) q}{1155 f_p r_p^2} , \\
     \PhiP_{4, 2, 1} &= \frac{\sqrt{\frac{\pi }{10}} \left(49 f_p^3-87 f_p^2+19 f_p-5\right) q}{462 f_p (3 f_p-1)
   r_p^2} , \\
     \PhiP_{4, 4, 1} &= -\frac{\sqrt{\frac{\pi }{70}} \left(3 f_p^2+4 f_p+1\right) q}{33 f_p r_p^2} , \\
     \PhiP_{6, 0, 1} &= -\frac{\sqrt{\frac{\pi }{13}} \left(f_p^2-6 f_p+25\right) q}{1232 f_p r_p^2} , \\
     \PhiP_{6, 2, 1} &= -\frac{\sqrt{\frac{\pi }{1365}} \left(f_p^2+2 f_p-63\right) q}{352 f_p r_p^2} , \\
     \PhiP_{6, 4, 1} &= \frac{\sqrt{\frac{\pi }{182}} \left(f_p^2-6 f_p-7\right) q}{176 f_p r_p^2} , \\
     \PhiP_{6, 6, 1} &= \frac{\sqrt{\frac{3 \pi }{1001}} (f_p+1)^2 q}{32 f_p r_p^2}.
\end{align}
\end{subequations}
\begin{widetext}
\begin{subequations}
\begin{align}
     \PhiP_{1, 1, 2} &= \frac{\sqrt{\frac{\pi }{6}} \left(690 f_p^4-1413 f_p^3+1039 f_p^2-283
   f_p+127\right) q}{3080 f_p^{3/2} (3 f_p-1) r_p^3} , \\
     \PhiP_{3, 1, 2} &= \frac{\sqrt{\frac{\pi }{21}} \left(31400 f_p^4-54651 f_p^3+32223 f_p^2+1119
   f_p-2851\right) q}{137280 f_p^{3/2} (3 f_p-1) r_p^3} , \\
     \PhiP_{3, 3, 2} &= \frac{\sqrt{\frac{\pi }{35}} \left(1128 f_p^4+799 f_p^3-3803 f_p^2-7043
   f_p+3343\right) q}{82368 f_p^{3/2} (3 f_p-1) r_p^3} , \\
     \PhiP_{5, 1, 2} &= -\frac{\sqrt{\frac{\pi }{330}} \left(1039 f_p^4-1464 f_p^3+816 f_p^2-564
   f_p+181\right) q}{4368 f_p^{3/2} (3 f_p-1) r_p^3} , \\
     \PhiP_{5, 3, 2} &= \frac{\sqrt{\frac{\pi }{385}} \left(561 f_p^4-1802 f_p^3+1318 f_p^2-266
   f_p+85\right) q}{3744 f_p^{3/2} (3 f_p-1) r_p^3} , \\
     \PhiP_{5, 5, 2} &= -\frac{\sqrt{\frac{\pi }{77}} \left(49 f_p^4+26 f_p^3-6 f_p^2+18
   f_p+1\right) q}{1248 f_p^{3/2} (3 f_p-1) r_p^3} , \\
     \PhiP_{7, 1, 2} &= \frac{\sqrt{\frac{\pi }{210}} \left(717 f_p^4-8460 f_p^3-5506 f_p^2+3156
   f_p+509\right) q}{155584 f_p^{3/2} (3 f_p-1) r_p^3} , \\
     \PhiP_{7, 3, 2} &= \frac{\sqrt{\frac{\pi }{70}} \left(109 f_p^4+9028 f_p^3-7794 f_p^2+2052
   f_p-691\right) q}{155584 f_p^{3/2} (3 f_p-1) r_p^3} , \\
     \PhiP_{7, 5, 2} &= -\frac{\sqrt{\frac{\pi }{770}} \left(527 f_p^4-260 f_p^3-246 f_p^2+348
   f_p-193\right) q}{14144 f_p^{3/2} (3 f_p-1) r_p^3},
\end{align}
\end{subequations}
\end{widetext}

\begin{align}
     \PhiP_{7, 7, 2} &= \frac{3 \sqrt{\frac{\pi }{1430}} (f_p+1)^2 (9 f_p+5) q}{1088 f_p^{3/2}
   r_p^3} , \tag{C4j} \\
     \PhiP_{9, 1, 2} &= \frac{\sqrt{\frac{5 \pi }{38}} \left(f_p^3-9 f_p^2+43 f_p-203\right) q}{155584
   f_p^{3/2} r_p^3}, \tag{C4k} \\
     \PhiP_{9, 3, 2} &= \frac{\sqrt{\frac{5 \pi }{4389}} \left(f_p^3-6 f_p^2+9 f_p+80\right) q}{7072
   f_p^{3/2} r_p^3}, \tag{C4l} \\
     \PhiP_{9, 5, 2} &= \frac{\sqrt{\frac{\pi }{2717}} \left(f_p^2-6 f_p-7\right) q}{544 f_p^{3/2}
   r_p^3}, \tag{C4m} \\
     \PhiP_{9, 7, 2} &= -\frac{\sqrt{\frac{5 \pi }{2717}} (f_p-7) (f_p+1)^2 q}{2176 f_p^{3/2}
   r_p^3}, \tag{C4n} \\
     \PhiP_{9, 9, 2} &= -\frac{\sqrt{\frac{5 \pi }{46189}} (f_p+1)^3 q}{128 f_p^{3/2} r_p^3}. \tag{C4o}
\end{align}\\

\endgroup

\bibliography{bibfile}
\end{document}